\documentstyle[prd,aps,floats,epsfig]{revtex} 
 
\newcommand{\Od}{{\cal O}}

\newcommand{\tr}{\mbox{tr}}

\newcommand{\im}{\mbox{Im}} 
\newcommand{\re}{\mbox{Re}}

\newcommand{\fpi}{f_\pi} 
\newcommand{\fpite}{f_\pi^t} 
\newcommand{\fpisp}{f_\pi^s} 
\newcommand{\ft}{f(t)} 
\newcommand{\fpit}{f_\pi(t)}

\newcommand{\ftsq}{f^2(t)} 
 
\newcommand{\fdot}{\dot f(t)} 
\newcommand{\fddot}{\ddot f(t)}

\newcommand{\intc}{\int_C dt \int d^3 \vec{x}} 
\newcommand{\intcc}{\int_C \! d^4x} 
\newcommand{\vxt}{(\vec{x},t)} 
 
\newcommand{\be}{\begin{equation}} 
\newcommand{\ee}{\end{equation}} 
\newcommand{\ba}{\begin{eqnarray}} 
\newcommand{\ea}{\end{eqnarray}} 
 
\newcommand{\NP}[1]{{Nucl.\ Phys.\ }{#1}} 
\newcommand{\ZP}[1]{{Z.\ Phys.\ }{#1}} 
 
\newcommand{\PL}[1]{{Phys.\ Lett.\ }{#1}} 
 
\newcommand{\AN}[1]{{Ann. Phys. } (N.Y.) {#1}} 
 
\newcommand{\PRep}[1]{{Phys.\ Rep.\ }{#1}} 
\newcommand{\PR}[1]{{Phys.\ Rev.\ }{#1}} 
\newcommand{\PRL}[1]{{Phys.\ Rev.\ Lett.\ }{#1}} 
 
\newcommand{\IJmp}[1]{{Int.\ J.\ Mod.\ Phys.\ }{#1}} 
 
\newcommand{\dnote}[1]{} 
\newcommand{\dla}[1] {\label{#1}} 
\newcommand{\dbib}[1] {\bibitem{#1}}

\begin{document} 
\draft 
\twocolumn[\hsize\textwidth\columnwidth\hsize\csname 
@twocolumnfalse\endcsname

\title{Pion production in 
nonequilibrium Chiral Perturbation Theory} 
 
\author{A. G\'omez Nicola} 
\address{Departamento de 
F\'{\i}sica Te\'orica II,  Universidad Complutense. 28040 
Madrid. SPAIN.} 
\date{\today} 
\maketitle

\begin{abstract} 
We apply the formalism  of Chiral Perturbation Theory out of thermal 
equilibrium to describe explosive production of pions via the 
parametric resonance mechanism. To lowest order the lagrangian is that 
of the Nonlinear Sigma Model where the pion decay constant becomes a 
time-dependent function. This model allows for a consistent 
nonequilibrium formulation within the framework of the 
closed time path 
method, where one-loop effects can be systematically accounted 
for and renormalized. We work in the narrow resonance regime where 
there is only one resonant band. The  pion 
distribution function is peaked around the resonant band 
 where the number of pions grow 
exponentially in time. The present approach is limited to remain 
below the back-reaction time, although it accounts for nearly all 
the pion production during the typical plasma lifetime. 
 Our results  agree with the analysis performed in the  $O(4)$ model. 
  The space and time components  $\fpi^{s,t} (t)$ 
are also analyzed. To one loop $\fpisp\neq \fpite$ unlike the 
equilibrium case and their final central 
values are  lower than the initial ones. 
This effect can be interpreted in terms of a reheating of the plasma. 
 \end{abstract} 
\vskip2pc] 
 
\section{Introduction}

Nonequilibrium Quantum Field Theory has attracted a considerable 
attention over the past decade, partially motivated by the 
experiments on  Relativistic Heavy Ion Collisions (RHIC). The 
ultimate goal of such experiments is  to describe the 
 properties of the QCD phase diagram and the Quark-Gluon Plasma (QGP) from 
 the observation of the final hadronic spectra. 
 The Relativistic Heavy Ion Collider at  BNL, which is already 
 running, will  reach  energy densities high enough to  confirm the 
 very promising results obtained at CERN (SPS) during the last few 
 years, indicating the existence of the QGP \cite{zsch01}. 
For a recent update of both theoretical and experimental results 
in this area see \cite{qm01}. 
A lot of theoretical effort has been put in trying to understand 
the various properties of  the  QCD phase diagram in different 
ranges of temperatures and densities in thermal equilibrium. 
  In addition, there are 
 many aspects of the nonequilibrium behaviour of the plasma which 
are  not fully understood. In the standard picture of the 
collision \cite{bjo83,lebellac} the plasma formed in the central 
rapidity region cools down very rapidly, reaching approximate local 
thermal equilibrium. During the subsequent  expansion, 
 the temperature scales of 
chiral phase transition and deconfinement are crossed and hadrons are 
produced. In this regime,  observables depend only on proper time 
approximately, in the central region. 
The expansion goes on until 
 the final freeze-out of hadrons. The typical plasma lifetime 
during which nonequilibrium effects are important and most of the 
final hadrons are produced is about 10 
 fm/c. 
 
 A  possible scenario to explain the observed final hadron distributions 
 is that where strong fluctuations of the pion 
field are formed during the chiral phase transition, giving rise 
to  the  so called Disoriented Chiral Condensates (DCC) 
\cite{an89}. These were suggested originally as misaligned vacuum 
 regions where the chiral field is pointing  out 
 in a different direction in isospin 
 space from that where the vacuum expectation value of the 
pion field vanishes. Ideally, one would have misaligned regions of 
observable size, with only neutral (or only charged) pions. 
 If such regions were formed, one would observe  large 
 clusters  of pions emitted coherently from the plasma as the pion 
 field relaxes to the normal vacuum \cite{rawi93}. This kind of 
 behaviour is 
indeed 
 observed in  Centauro  events  in 
 cosmic ray experiments \cite{centauro}. 
 A clear signal for DCC formation has not yet been observed in 
 RHIC experiments   \cite{fermilab}.  However, it seems that one has to 
 measure higher order pion correlation functions in order to 
 identify a pure DCC signal which is not masked by other effects 
 \cite{hiro,ble00} so that  the search will continue at 
BNL. Other observable consequences of DCC-like 
 configurations would be an enhancement of dilepton and photon 
 production \cite{cooper99,boya97} and a modification of the 
 effective $\pi^0\rightarrow\gamma\gamma$ vertex \cite{prem00}. 
 In any case, pion production within the energy scales of 
 the chiral symmetry provides a natural framework for 
 hadronization \cite{mm95}.

Thus, one should be able to describe nonequilibrium phenomena such 
as large pion production,  from  the microscopic theory governing 
the relevant degrees of freedom. At the energy scales where the 
chiral symmetry plays a predominant role (below 1 GeV) QCD is 
nonperturbative and one has to use an effective lagrangian which 
describes satisfactorily the microscopic 
 meson dynamics.  Such a  theory must incorporate  the QCD 
 symmetries 
and    the chiral spontaneous symmetry breaking (SSB) pattern. In 
this picture the Nambu-Goldstone bosons (NGB) are the lightest 
mesons ($\pi$, $K$, $\eta$) and the masses of the light quarks are 
meant to be treated perturbatively. One possible choice is simply 
the 
 $O(N)$  model where the fundamental fields are  $N-1$ pions and 
the $\sigma$, 
 and the potential has  the typical SSB shape. 
 However, one should bear  in mind that this model becomes 
nonperturbative in the coupling constant at low energies so that 
it is imperative to perform alternative  expansions such as  large 
$N$. 
 On the other hand, the $O(N)$ model   shares 
 the QCD chiral symmetry breaking 
 pattern  only for $N=4$, so that it is not able to incorporate kaons 
 and etas. 
 
 An alternative approach  is an effective theory built as an 
 infinite sum of terms with increasing number of derivatives, 
only in terms of the  NGB fields. 
 The   Nonlinear Sigma Model 
 (NLSM) is the lowest order action one can write down in this 
 expansion. Higher order   corrections   come both from  NGB loops and 
 higher order lagrangians and  can be  renormalized 
 order by order in energies, yielding  finite predictions 
 for  the meson observables. The unknown 
 coefficients, which  encode all the information on the 
 underlying theory,  absorb the loop infinities and their finite 
 part 
 can be fitted to experiment. 
 This framework constitutes the so called chiral perturbation theory (ChPT) 
 \cite{we79,gale} which provides a well-defined   perturbative 
 expansion in 
terms 
of $p/\Lambda_{\chi}$, 
  where  $p$ stands generically for any  meson 
 energy scale of  the theory 
(masses, external momenta, temperature 
 and so on) and 
  the  chiral scale $\Lambda_{\chi}\simeq$ 1.2 GeV 
(see \cite{dogoho92,dogolope97} for a 
 review). One of the many advantages of this scheme is that it can 
 be extended from the $SU(2)$ chiral symmetry (only $\pi$ fields) to 
 $SU(3)$ with $K$ and $\eta$. The ChPT formalism has also been 
 applied  in  thermal equilibrium  to 
 analyse various properties   of the low temperature 
 meson gas  \cite{gale87,gele89,boka96}.

 In 
the context of nonequilibrium chiral dynamics, two possible 
scenarios for pion production and DCC 
 formation have been 
proposed: the first one takes place in the early stages of the 
plasma evolution. Roughly speaking, after a very rapid cooling 
 the chiral field is at the top of the classical 
potential in the chirally broken phase. As the field rolls down, 
long wavelength modes grow exponentially (spinodal instabilities) 
and  this behaviour is responsible for the enhancement of DCC's. 
There have been several approaches in the literature to implement 
this idea in the $O(4)$ model 
 \cite{rawi93,bodeho95,cooper9596}. 
 Typically, the pion distribution function is peaked at low 
 momenta, being  different from a thermal distribution 
 \cite{cooper9596} whereas  the pion 
densities and DCC sizes predicted are around  $n_\pi\simeq$ 0.2 
fm$^{-3}$ and  1.5-2 fm respectively. 
The second suggestion is based on the parametric resonance 
mechanism \cite{mm95} and inherits the idea from inflationary 
reheating \cite{linde}.  The analysis in the spinodal regime shows that 
the time it takes for the field to roll 
down to the bottom is very short compared with the total plasma  lifetime. 
Thus, in the parametric resonance approach, the 
$\sigma$ field is {\em oscillating} around the minimum of the 
potential   in a later stage of the plasma evolution. Those 
oscillations  transfer energy to the pion modes, giving rise to 
pion solutions exponentially growing in time via parametric 
resonance. Typically, the unstable modes develop in bands in 
momentum space and the more important resonance 
 band is centered at  $k\simeq 
m_{\sigma}/2$ \cite{hiro,randrup00}. 
The DCC sizes in this approach can be as 
large as 5 fm \cite{kaiser} and recent calculations 
show that  strong charge-neutral 
correlations in parametric resonance can 
 be used to identify a pure DCC signal \cite{hiro}. 
 Furthermore, the reheating 
process yields predictions for the final hadronization temperature 
compatible with the observations \cite{mm95}. 
One must stress that both approaches are complementary and in fact 
the initial conditions needed for the parametric resonance 
correspond to the final stage of the rolling down solution. A very 
detailed analysis of both regimes  in the context of the $O(N)$ 
model can be found in \cite{boyaetal96}.

The purpose of this work is to explore pion production in 
parametric resonance within the ChPT framework, as a complementary 
analysis to the $O(4)$ model. We will 
 show that within  this formalism one 
can also describe regions where the number of pions and the pion 
correlator grow exponentially. The main advantages of the ChPT 
approach are that one can follow a consistent perturbative 
treatment which is renormalizable order by order and that 
 it can be extended to three flavours. Besides, one is dealing 
 only with NGB fields, although  we will show how the 
 pion production can be understood in terms of the 
$\sigma$ field evolution in the $O(4)$ model. 
 This method is  best 
 suited for the  stage of the plasma expansion 
 where the system is well into the broken phase of the chiral 
 symmetry.  This is precisely the regime where parametric resonance 
 takes place. 
 
We will build on a 
previous work \cite{agg99}, where we have analysed 
 the extension of ChPT  out of thermal 
equilibrium.  In that work it has been shown that the  power 
 counting and  renormalization program can be consistently 
implemented also at nonequilibrium. In turn, 
the present analysis 
 will provide 
a particular example where it will be shown explicitly how the chiral 
 power counting and renormalization program work, yielding predictions 
  for physical observables. The key idea  is to make use of the 
 derivative expansion consistently  implemented  in ChPT in order to 
 study the system not far from equilibrium. 
 For that purpose, the nonequilibrium dynamics is encoded 
effectively 
 in the parameters of the model. To leading order and assuming 
 a spatially homogeneous system, we let the pion decay constant be 
 time dependent. This function acts as an external force on the 
 pion degrees of freedom. It is important to bear in mind that 
 a self-consistent treatment should amount to 
 incorporate the full hydrodynamics  of both the fluid and 
 pion modes \cite{son00}. In the present approach we will concentrate only on 
 the influence of the expansion on the meson dynamics. This is a 
 similar situation as considering a QFT in the 
 presence of an external curved background space-time \cite{bida82}. 
In that case, it 
 makes sense under certain conditions to ignore the back-reaction 
 effect of the matter fields in the metric. 
 Similarly, we will see that for the time scales relevant for pion 
 production it is reasonable to ignore those effects and treat the 
 influence of the expanding plasma as external.

 The paper is organized as follows: In section \ref{model} we 
will review the nonequilibrium ChPT and its relationship to curved 
space-time QFT, which will be crucial in what 
follows. The parametric resonance approach in its simpler version 
will be discussed in section \ref{parres}, while sections 
\ref{fpi} and \ref{number} will be devoted to analyse the effects 
of parametric resonance in two different observables: the pion 
decay functions and the pion number respectively. 
 The latter is the most relevant observable as far as pion 
 production is concerned whereas the former will allow us to 
 estimate the time scale when the back-reaction effects become 
  important as well as the final temperature by that time. 
  In both cases we will calculate up to one-loop in ChPT, paying 
   particular attention to renormalization. Besides, in section 
   \ref{number}  the definition of particle number 
   we will use  and its relationship with the energy-momentum 
   tensor are discussed. Our conclusions are summarized in section 
\ref{conc}. 
We have included two appendices collecting  some useful results 
about the Mathieu equation and curved space-time QFT respectively. 
 
\section{Nonequilibrium Chiral Perturbation Theory} 
\dla{model} 
\subsection{The NLSM  out of thermal equilibrium} 
 
The chiral lagrangian to lowest order  is the nonlinear sigma 
model (NLSM), which contains two derivatives of the fields. In the 
chiral limit, where the mass of the light quarks is set to zero, 
the NLSM only contains one energy parameter $f$ . To lowest order, 
$f$ is nothing but the pion decay constant $f=\fpi\simeq$ 93 MeV 
($f\neq \fpi$ to higher orders). Consistently, our nonequilibrium 
model will be the NLSM where $f$ is replaced by a time-dependent 
function $\ft$. This function acts as an external field encoding 
the time evolution of the system as, for instance, the expansion 
of the plasma formed after a RHIC in proper time. 
As for the initial conditions we will assume that the system is in 
thermal equilibrium before some initial time ($t=0$ for 
convenience) at the temperature $T_i=\beta_i^{-1}$. This is an 
important simplification from the point of view of the 
nonequilibrium path-integral formulation. In fact, one can 
formulate the generating functional for the real-time Green 
functions by extending the time arguments to the contour $C$ in 
the complex plane showed in Figure \ref{contour}. This is an extension 
of the Closed Time Path  technique \cite{schkel}  for 
nonequilibrium field theory, where the inclusion of the 
imaginary-time leg is a consequence of the choice of equilibrium 
initial conditions \cite{boleesi93}. 
Thus, $f(t\leq 0)=f$, and  the function $\ft$ represents an 
external force switched on at $t=0$ and driving the system out of 
equilibrium. Note that we choose that departure to be 
instantaneous and then 
 $\ft$ cannot be analytical at $t=0$. 
 Nevertheless, it is natural to expect that these   restrictions on 
the initial conditions do not influence very much the physical 
results for longer times, especially if the system is not far from 
 equilibrium. 
 \begin{figure} 
\begin{center} 
\vspace*{-5cm} \hspace*{-1cm} 
\hbox{\epsfig{figure=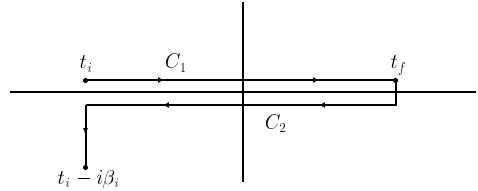,width=9.5cm}} 
\end{center} 
\vspace*{-5.5cm} \caption{The contour $C$ in complex time $t$. The 
lines $C_1$ and $C_2$ run  between $t_i+i\epsilon$ and 
 $t_f+i\epsilon$ and  $t_f-i\epsilon$ and $t_i-i\epsilon$ 
respectively, 
 with  $\epsilon\rightarrow 0^+$.} 
\dla{contour} 
\end{figure} 
 
 
  Thus, our starting point will be the following nonequilibrium NLSM 
  action: 
\begin{equation} 
S_2[U]=\intcc \ \frac{f^2(t)}{4} \ \tr \ \partial_\mu 
U^\dagger\vxt 
\partial^\mu U\vxt 
\dla{nlsmne} 
\end{equation} 
where $\intcc=\intc$.  We will restrict here to the case of two 
light quark flavors (i.e, the NGB are only the pions) and hence 
$U(x)\in 
 SU(2)$. Besides, we will be 
 interested only in the  chiral limit  i.e,  massless pions.  Thus, 
 we are not including any {\it explicit} symmetry-breaking 
 term in the action. Note that the action (\ref{nlsmne}) is chiral invariant 
($U\rightarrow LUR^\dagger$ where $L,R$ are constant $SU(2)$ 
matrices) by construction, which will play an important role in 
what follows. As it is customary, $U(x)$ is  parametrized in terms 
of the pion fields $\pi^a\vxt$ as: 
\ba U\vxt=\frac{1}{f(t)}\left\{\left[f^2(t)-\pi^2 \vxt 
\right]^{1/2} I +i\tau_a\pi^a\vxt\right\}  \nonumber\ea where $I$ 
is the $2\times 2$ identity matrix, $\tau_a$ are the Pauli 
matrices, $\pi^2=\pi^a \pi_a$  and 
$\pi^a(t_i-i\beta_i)=\pi^a(t_i)$ is the equilibrium (KMS) boundary 
condition, with $t_i<0$. As we will show  below, the physics does 
not depend  on the choice of $t_i$.

 It is useful to recall that the NLSM can be viewed {\it to 
lowest order} as an $O(4)$ model subject to the constraint 
$\sigma^2+\pi^2=f^2$. The same applies to the nonequilibrium case 
and therefore we can think of $\ft$ as the lowest order 
expectation value of the $\sigma$ field in the $O(4)$ model, which 
has not reached its equilibrium value yet. 
 
With the NLSM one can predict the very low energy behaviour of 
pion observables such as the pion decay constant or the pion 
scattering amplitude, which agree with the current 
algebra predictions. However, to go beyond the lowest order, one has to 
consider  pion loops. Thanks to Weinberg power counting theorem 
\cite{we79} we know that the loop diagrams are of the same order 
as lagrangians with more derivatives of the pion fields. In fact, 
it can be shown explicitly that the (undetermined) coefficients of 
such lagrangians absorb all possible UV divergences coming from 
the loops and hence one gets finite and scale-independent 
predictions for the pion observables. 
 
In the present  case, we need an extra ingredient to the power 
counting, namely,  the time derivatives of the 
function $\ft$. We will take 
 \begin{equation} 
\frac{\fdot}{\ft}\simeq \Od (p), \qquad \frac{\fddot}{\ft}, 
\frac{[\fdot]^2}{f^2(t)} \simeq\Od (p^2) \dla{chipoco} 
\end{equation} 
and so on. In this sense,  one remains close to equilibrium. 
The rest of the power counting 
is the usual one, i.e,  every derivative of the pion field is $\Od 
(p)$ and  every pion loop introduces an extra $\Od(p^2)$. Other 
than being subject to the conditions (\ref{chipoco}), we will let 
$\ft$ be arbitrary. However, in this work we shall discuss how 
$\ft$ can be chosen 
consistently with physical results such as pion production.

Let us now expand  the NLSM action to lowest order in the pion 
fields. Using the boundary conditions, we can  integrate by parts 
and write, to second order in the pion fields, 
\ba 
S_2[\pi]=- \frac{1}{2}\intcc \pi^a\vxt\left[\Box + m^2 
(t)\right] \pi^a\vxt\ + \dots\dla{act2pi} 
\ea 
where $m^2(t)=-\fddot/\ft$. That is, the model accommodates a 
time-dependent pion mass term, {\it without} breaking explicitly 
the chiral symmetry (unlike a physical pion mass term). This is an 
important feature of this model, since it suggests an interesting 
connection with a QFT in the presence of an 
external curved space-time background. We will discuss this point 
in detail in  section \ref{rencur}. Before that, let us analyse 
the main properties of the leading order two-point function (the 
propagator) arising from the above NLSM action.

\subsection{The leading order propagator} 
\dla{lop} 
 
The two-pion correlation function is 
$i G^{ab}(x,x')= \langle T_C \pi^a (\vec{x},t) \pi^b 
(\vec{x'},t') \rangle$, where $T_C$ indicates time ordering along 
the contour $C$. Note that, from isospin invariance, we can write 
just $G^{ab}(x,x')=\delta^{ab} G(x,x')$. Besides,  by spatial 
translation invariance  the two-point function depends only on 
$\vec{x}-\vec{x}'$. This is not true for the time coordinates 
 due to the nonequilibrium time dependence in the 
lagrangian. 
 
There are four different types of propagator depending on the 
relative position of $t$ and $t'$, namely, $G^{11}$, $G^{12}$ and 
so on \cite{lebellac}. Instead of writing all the combinations 
explicitly, we shall keep the condensed notation of time ordering 
with  $T_C$ defining the natural extensions 
$\theta_C$, 
$\delta_C$ and so on. Thus, from (\ref{act2pi}), the leading 
order  $G_0 (x,x')$ is a solution of the differential equation 
\begin{equation} 
\left\{\partial_t^2 + k^2 + m^2(t)\right\} G_0 (t,t',k)=-\delta_C 
(t-t') \dla{lopropeq} 
\end{equation} 
where $G_0 (t,t',k)$ is the Fourier transform in the space 
coordinates only  and $k^2=\vert \vec{k} \vert^2$. 
 
As for the boundary conditions, thermal equilibrium for $t<0$ 
means that we have to impose KMS boundary conditions at the 
imaginary-time leg in Figure \ref{contour} \cite{lebellac}. That 
is, defining \ba G(t,t',k)&=&G^> (t,t',k) \theta_C 
(t-t')\nonumber\\ &+& G^< (t,t',k) \theta_C (t'-t) 
,\dla{adretdef}\ea the KMS boundary conditions  read 
\be 
G_0^> (t_i-i\beta_i,t',k)=G_0^< (t_i,t',k) \dla{KMS} \ee 
 
The general solution to the differential equation (\ref{lopropeq}) 
with the boundary conditions (\ref{KMS}) can be constructed for 
all the branches of the contour in terms of two particular 
solutions $h_{1,2}(t,k)$ to the homogeneous equation \cite{sewe85}: 
\be 
\left\{\partial_t^2 + k^2 + m^2(t)\right\} h_i (t,k)=0 \qquad 
i=1,2, \dla{homo} \ee  such that their 
Wronskian 
\be 
W(t,k)\equiv \dot h_1 (t,k) h_2 (t,k)-h_1 (t,k) \dot h_2 (t,k)\neq 
0 \ee

It is important to remark that the general solution $G_0(t,t',k)$ 
must be continuous and differentiable in the time coordinates so 
that it is uniquely defined. Thus, in our case we 
demand that $h_i (t,k)$ for $t>0$ and their first derivative match 
the equilibrium solutions at $t=0$. Since $m^2(t<0)=0$, two 
independent equilibrium solutions are  given for $t<0$ by 
\be 
h_1^{eq} (t,k) = \frac{i}{\sqrt{2k}} e^{-i kt}\quad ; \quad 
h_2^{eq} (t,k) = \frac{1}{\sqrt{2k}} e^{i kt},\dla{twoeqsol}\ee 
which we have normalized so that $W^{eq}(t,k)=1$. It is not 
difficult to see that continuity and differentiability 
 at $t=0$ imply that $W(t,k)=1$ also for $t>0$. On 
the other hand, since $m^2(t)$ is real, if $h_1 (t)$ is a solution 
to (\ref{homo}), so it is $h_2(t)=ih_1^* (t)$, where $W(t)=1$ if 
it matches the equilibrium solution at $t=0$.

We shall be dealing here with  real time evolution for positive 
time coordinates and therefore, unless otherwise stated, we will 
be interested in $G^{11}$ only. In that case, we will suppress the 
"11" superscript for simplicity.  Nonetheless, it should be borne 
in mind that in the loop integrals there are contributions from 
all the branches of the contour \cite{agg99}.

Thus, the solution for $G_0^{11}$ is given by \cite{sewe85} \ba i 
G_0 (t,t',k) &=&  h_1(t,k) h_1^* (t',k) \theta (t-t')\nonumber\\ 
&+& h_1^* (t,k) h_1 (t',k) \theta (t'-t)\nonumber\\ 
 &+&  n_B (k) \left[h_1(t,k) h_1^* (t',k) +  h_1^* (t,k) h_1 (t',k) 
 \right] 
\nonumber\\ 
\dla{losol}\ea where $t$ and $t'$ are both positive and the 
boundary conditions at $t=0$ imply 
\be 
h_1(t=0^+,k)=\frac{i}{\sqrt{2k}} \quad ; \quad \dot 
h_1(t=0^+,k)=\sqrt{\frac{k}{2}} \dla{ic}\ee where the dot means 
$d/dt$. The dependence with the initial temperature appears 
through the Bose-Einstein distribution function 
\be 
 n_B (k)=\frac{1}{e^{\beta_i k}-1} 
\dla{be} \ee

In different parts of this work we will need the two-point 
function evaluated at the same space-time points: \ba 
G_0(t)&\equiv& G_0 (x,x)= \int \frac{d^{d-1} k}{(2\pi)^{d-1}} G_0 
(t,t,k)\nonumber\\&=&\frac{2}{\Gamma (\frac{d-1}{2}) 
(4\pi)^{\frac{d-1}{2}}} \int_0^\infty dk k^{d-2}  G_0 (t,t,k) 
\dla{g0def} \ea where  the equal-time correlator in momentum space 
reads, from (\ref{losol}): 
\be 
 i G_0 (t,t,k)=\left[1+2 n_B (k)\right] \ \vert h_1 (t,k) \vert^2 
\dla{g0tk}\ee

 Because 
of the loss of time translation invariance, $G_0(t)$  is a 
time-dependent quantity. Besides, it may be  UV divergent  and, 
therefore, we will use dimensional regularization (DR) with $d$ 
the space-time dimension, which is a suitable regularization 
scheme as far as chiral lagrangians are concerned 
\cite{gale,dogoho92,dogolope97}.

\subsection{Renormalization and curved space-time} 
\dla{rencur} 
 
Once we have defined our nonequilibrium power counting, we can 
apply 
 ChPT to calculate the time evolution of the observables. 
 In doing so, we must pay special attention to renormalization. 
The fact that there is a time-dependent mass term indicates that 
there can be new time-dependent infinities in the chiral loops. 
For instance, in standard ChPT with a nonzero pion mass, the 
tadpoles renormalizing the pion propagator to lowest order yield 
the usual infinities proportional to $m_\pi^2$ in DR. These 
infinities are absorbed by two counterterms 
 proportional to $m_\pi^2$ in the fourth order  lagrangian 
\cite{gale}. We expect similar divergent contributions here 
proportional to $m^2 (t)$. However, we are working in the chiral 
limit and therefore  we are not allowed to introduce the above 
mentioned counterterms. Otherwise we would break explicitly the 
chiral symmetry. Hence, we should be able to construct the most 
general fourth order action, which in particular has to include 
new terms (and hence new low-energy constants) to cancel those 
extra divergences, preserving exactly the chiral symmetry. 
 
There is a natural way to  find this $\Od(p^4)$ lagrangian, 
 using  a very fruitful analogy: the action 
 (\ref{nlsmne}) can be written as a 
   NLSM on a curved space-time background 
 corresponding to a spatially flat Robertson-Walker (RW) metric, 
 with scale factor 
  $a(t)=f(t)/f$. For that purpose  it is more convenient to work 
   in terms of rescaled fields  $\tilde \pi^a 
  \vxt=\pi^a\vxt/a(t)$. Hence, 
 $U=((f^2-\tilde\pi^2)^{1/2} I + i\tau_a \tilde \pi^a)/f$ and 
 we can write the action (\ref{nlsmne}) as 
\begin{equation} 
S_2[U]=\intcc \ \frac{f^2}{4} \ \sqrt{-g} g^{\mu\nu} 
 \tr \ \partial_\mu U^\dagger\vxt 
\partial_\nu U\vxt 
\dla{Scurv} 
\end{equation} 
where   the metric $g^{\mu\nu}$ is nothing but the spatially flat RW 
metric with  line element 
 $ds^2=a^2(t)[dt^2-d\vec{x}^2]$ in conformal time $t$ \cite{bida82} 
 and   $g\equiv\det g=-a^{8} (t)$. 

With our chiral power counting, it is straightforward to assign 
the chiral order of the covariant tensors constructed from the 
metric. For instance, the Ricci tensor $R_{\mu\nu}$ is $\Od (p^2)$ 
and so is the Ricci scalar $R=g^{\mu\nu} R_{\mu\nu}$, and so on. 
Explicit expressions for these tensors and other useful results 
for this metric are collected in Appendix \ref{ap:curved}. An 
important point in this formulation is that we are considering the 
so called minimal coupling of the matter fields with the metric. 
That is, we are discarding possible couplings  between the pion 
fields 
 and $R(x)$ to 
$\Od (p^2)$, such as $\xi R(x) \tr (U+U^\dagger)$. The reason is 
 that we want to preserve chiral invariance, which would 
be broken by those terms \cite{dole91}. Thus, in this language, 
 $m^2(t)$ in (\ref{act2pi}) represents  the minimal coupling with the RW 
metric preserving chiral 
 invariance.

Therefore, we have a systematic way to construct the 
nonequilibrium lagrangian to any order. We just have to include 
all possible terms consistent with the chiral symmetry, 
contracting indices covariantly with the metric $g_{\mu\nu} (x)$. 
In particular, to $\Od(p^4)$ it reads \cite{dole91}: 
\ba 
S_4 [U,g,R]&=&\intcc \sqrt{-g}\left[ {\cal 
L}_4[U,g]\right.\nonumber\\&-&\left.\left(L_{11}R 
g^{\mu\nu}+L_{12}R^{\mu\nu}\right) \tr 
\partial_\mu U^\dagger 
\partial_\nu U\right] 
\dla{lag4gen} \ea where ${\cal L}_4[U,g]$ stands for  the standard 
(equilibrium) lagrangian \cite{gale} 
 with indices raised and lowered with the $g^{\mu\nu}$ metric and 
the rest are new $\Od(p^4)$ invariant couplings with   $R(x)$ and 
$R_{\mu\nu} (x)$ in the chiral limit. These are the new terms we 
need, where 
 $L_{11}$ and $L_{12}$ are the new coupling constants. 
 We are following the notation of 
 \cite{dole91}, where this lagrangian was first considered to study the 
energy-momentum tensor of QCD at low energies \footnote{The terms 
with $L_{11}$ and $L_{12}$ in (\ref{lag4gen}) differ in a global 
sign from those in \cite{dole91}. The reason is that we are 
following here the convention in \cite{weinberg} for the Riemann 
tensor, namely 
$R_{\beta\gamma\delta}^\alpha=\partial_{\delta} 
\Gamma_{\beta\gamma}^{\alpha}-\dots$, 
where $\Gamma_{\beta\gamma}^{\alpha}$ are the Christoffel symbols 
(see Appendix \ref{ap:curved}) whereas in \cite{dole91} the 
convention for $R_{\beta\gamma\delta}^\alpha$ is reversed in sign. 
 For the same reason, every term 
 proportional to $L_{11}$ or $L_{12}$ here has its sign changed 
  with respect to those in \cite{agg99}.}. 
  The same lagrangian has been used in the context of pion hard 
  exclusive production \cite{leh00}. In 
\cite{dole91} it has been shown  that in order to cancel the 
one-loop infinities, $L_{11}$ has to be renormalized as 
\be 
L_{11}=L_{11}^r (\mu) + \frac{1}{6}\frac{\mu^{d-4}}{16\pi^2}\left[ 
\frac{1}{d-4}-\frac{1}{2}\left(\log 4\pi - \gamma +1\right)\right] 
\dla{l11ren}\ee with $d$  the space-time dimension, $\gamma$ the 
Euler constant and $\mu$ the renormalization scale. $L_{11}^R 
(\mu)$ is finite and depends on $\mu$  so that the combination in 
the r.h.s of (\ref{l11ren}) remains scale-independent. On the 
other hand, $L_{12}$ is finite. Their numerical  values can be 
obtained from the experimental information on  the QCD 
energy-momentum form factors. They yield  $L_{12}\simeq -2.7\times 
10^{-3}$ and $L_{11}^r(\mu=1$ GeV$)\simeq 1.4\times 10^{-3}$. We 
will use these same values here since their possible 
nonequilibrium corrections are of higher order in our analysis.

Thus in our case we have to replace in (\ref{lag4gen}) the RW 
metric. After expanding in the $\pi^a$ fields and 
partial integration we have: 
\ba S_4[\pi,g]&=& - \frac{1}{2}\intcc\pi^a\left[f_1 
(t)\partial_t^2-f_2 (t)\nabla^2 
 + m_1^2 (t)\right]\pi^a \nonumber\\&+&\Od(\pi^4) 
\dla{lag4} \ea with 
\begin{eqnarray} 
f_1(t)&=&-12\left[\left(2L_{11}+L_{12}\right)\frac{\fddot}{f^3(t)} 
-L_{12}\frac{[\fdot]^2}{f^4 (t)}\right] \nonumber\\ 
f_2(t)&=&-4\left[\left(6L_{11}+L_{12}\right)\frac{\fddot}{f^3(t)} 
+L_{12}\frac{[\fdot]^2}{f^4 (t)}\right] \nonumber\\ m_1^2 
(t)&=&-\left[\frac{f_1(t)\fddot+\dot f_1 (t)\fdot}{\ft}+ 
\frac{1}{2}\ddot f_1 (t)\right] \dla{f12} 
\end{eqnarray}

The above lagrangian should take care of the nonequilibrium 
infinities we might find in the pion two-point function. As far as 
this work is concerned, these are the only infinities we will have to 
renormalize. 
 
In the following sections we will concentrate on a particular case 
for  $\ft$ (or the scale factor if we use the curved space time 
terminology) which is of physical relevance as the simplest 
approximation producing a large number of correlated pions. In 
addition, this example will allow us to test explicitly the 
cancellation of the (new) nonequilibrium infinities appearing in 
the observables considered.

\section{Parametric resonance and pion production} 
\dla{parres} 
 
\subsection{The parametric resonance approach} 
It is clear that  our approach will be useful in a stage of the 
plasma evolution when the departure from equilibrium is of the 
same order as the meson energies. Hence, as far as pion production 
is concerned, we are in  the  parametric resonance regime.  Let us 
briefly review some of the ideas behind parametric resonance 
 in the $O(4)$ model \cite{mm95,boyaetal96,hiro}. 
 In the last stage of the field evolution, the 
$\sigma$ field is oscillating near the true vacuum and those 
oscillations have relatively small amplitude \cite{cooper9596}. 
 Following  a semiclassical approach, 
  one can split the $\sigma$ field  as 
$\sigma (\vec{x},t)=\sigma_0 (t)+\delta\sigma(\vec{x},t)$ where 
 $\sigma_0 (t)$  is  a {\it time-dependent} homogeneous 
classical background, solution of the equations of motion to 
leading order in the amplitude, whereas $\delta\sigma$ includes 
next to leading order  corrections and quantum fluctuations. One 
can proceed perturbatively around the classical solution $\sigma_0 
(t)$. In a first approximation, quantum fluctuations of both the 
$\sigma$ and the pions can be neglected. Thus, if the $\sigma$ 
field is oscillating around the potential minimum $\sigma=f$ and 
the amplitude of the oscillations $q$ is small, one can solve the 
equations of motion perturbatively in $q$. To leading order the 
equations of motion for the $\sigma$ and the pions decouple from 
each other and one simply gets $\sigma_0(t)/f=1-(q/2) [\sin( 
m_\sigma t+\varphi)-\sin\varphi]$ where $m_\sigma$ is the $\sigma$ 
mass, $\varphi$ is an arbitrary phase and we have chosen the 
initial conditions so that the field is at the bottom of the 
potential for $t=0$, consistently with our choice of the initial 
equilibrium state. Neglecting the pion correlations is equivalent 
to state that $\langle \pi^2 \rangle \ll f^2$. 
 If the first order solution is inserted into the equations of 
 motion to next to leading order, the pions satisfy a 
 Mathieu equation. The importance of this equation is that 
  it  has solutions 
 exponentially growing in time for certain bands in momentum 
 space. This is the essence of the parametric resonance mechanism, 
 which even in this simple classical picture is consistent with 
  hadronization  \cite{mm95}. 
 The parametric resonance idea is directly imported from reheating 
and preheating in inflationary cosmology \cite{linde}, where the 
small $q$ approach is called the narrow resonance limit for 
reasons to become clear below.

The approach described in the previous paragraph is the crudest 
one can follow in this context, although it reproduces the main 
features of parametric resonance. One can refine it in several 
ways. First, neglecting the pion correlations but keeping the NLO 
terms in the $\sigma$ amplitude leads to a Lam\'e equation instead 
of the Mathieu equation \cite{boyaetal96,kaiser}. As it is 
emphasized in \cite{boyaetal96}, the difference is not only 
quantitative, but the resonance structure is also different. As 
for the importance of pion correlations, one must bear in mind 
that this is a matter of time scales. Such correlations grow 
exponentially in time until eventually they reach the same order 
of magnitude as the $\sigma$ term. As it is customary, we will 
refer to that time scale as the back-reaction time, so that, 
typically $\langle \pi^2 \rangle (t_{BR})\sim f^2$.  The name is 
again inherited from Cosmology where the back-reaction describes 
the modifications of the metric or the inflaton field due to 
quantum fluctuations of the matter fields \cite{linde,bida82}. It 
is clear that for $t\geq t_{BR}$, pion correlations must be 
included self-consistently in the dynamics of the $\sigma$ field. 
In $O(N)$ models this usually requires numerical 
 simulations. For instance, in the large $N$ limit it has been shown 
 that, when the back-reaction is properly accounted for, 
  pion correlations are responsible for the {\it damping} 
 of the $\sigma$ field from $t_{BR}$ onwards and, more importantly, 
  this dissipation  stops the exponential growth in the pion number 
\cite{boyaetal96,boyaetal95}. 
   A different story  though is that one can interpret that process as 
   thermalization, or in other words,  that the final 
   particle spectrum is thermal. In fact, that is not  the case 
  when pion amplification occurs \cite{cooper9596}. 
  Thus, in practice, all the interesting physics associated with pion 
 production in parametric resonance takes place before the 
 back-reaction time. 
  Finally, a word must be said about  the quantum corrections 
  of the $\sigma$ field, which can be treated semiclassically \cite{hiro}. 
  In the narrow resonance approach, 
 the evolution of the pion fluctuations is influenced only by 
 terms linear in $\sigma_0 (t)$ and the dynamics of the $\sigma$ 
 fluctuations is not important for pion production, since the 
 width of the resonance band for the $\sigma$ is 
 negligible with respect to that of the pions. Thus, the essence 
 of the exponential growing of pion fluctuations is not changed 
 qualitatively by including quantum corrections in $\delta\sigma$.

At this point, let us establish the connection with our present 
ChPT approach. Our philosophy will be to work out consistently 
 the simplest choice for $\ft$ yielding parametric 
resonance. For that purpose, it is useful to compare with the 
 $O(4)$ model.  In the oscillatory regime 
(with small oscillations)  one starts by  keeping only the leading 
order $\sigma\sim\sigma_0 (t)$. In that limit and for small 
oscillations  the NLSM is nothing but the $O(4)$ model subject to 
$\pi^2+\sigma^2=\sigma_0^2(t)$. In other words, we should take 
simply 
\be 
f(t)=f\left[1-\frac{q}{2}\sin Mt \right] \qquad (t>0) \dla{ft} 
 \ee 
where we have chosen  $\varphi=0$ so that 
$$m^2(t)=-(qM^2/2)\sin 
Mt+\Od(q^2).$$ Note that with this choice, not only $f(0^+)=f$ but 
also $m^2(0^+)=0$, i.e, both $\ft$ and $m^2(t)$ match their 
equilibrium values at $t=0$. This will play an important role in 
the analysis in section \ref{sec:drlo}. 

 Thus,  we would use the NLSM model (\ref{nlsmne}) with $\ft$ in 
 (\ref{ft}) if we were interested in describing pions 
out of equilibrium classically, for times where the back reaction 
is not important and the plasma is in the broken phase. 
Then,  we can calculate pion observables using ChPT, 
  where the pion fluctuations can be treated quantum mechanically 
  in a consistent fashion. As a matter of fact, 
   there is no need to invoke the $O(4)$ model in the first place, since 
   the NLSM is the lowest order action compatible with all the 
   symmetries, driven out of equilibrium through the time 
   dependence in $\ft$.  For that  reason, 
   we have replaced the mass of the sigma by $M$, an 
   arbitrary mass parameter. We expect it to be in a range 
   compatible with $m_\sigma$ but the advantage of our present 
   approach is that we do not have to worry about the 
    uncertainties related  to the $\sigma$ particle 
    \footnote{The $\sigma$ mass is not very well determined. 
    It ranges between 400-1200 MeV according 
   to the latest PDG data \cite{rpp2000}. It is not even clear that one 
    can describe it as a particle. For instance, it shows up in  pion-pion 
     scattering in ChPT as a rather broad resonance in the $I=J=0$ channel 
     \cite{olleretal}.}. 
     We will discuss below what range of 
   values for $M$ is reasonable in order to reproduce the right 
   order of magnitude in pion production. 
 
 Let us be more precise now about the smallness of  $q$ in our 
approach. According to our previous discussion about the 
nonequilibrium chiral power counting, we should demand at least 
that  $q M^2=\Od (p^2)$ 
and so on. In this way, all   the $\Od(p^4)$ corrections will 
remain under control, as we will see  below. It is 
important to remark that in this work we shall restrict to one 
loop in ChPT. Going beyond that would require additional 
restrictions on the value of $M$. 
Before carrying on, we would like to summarize the assumptions and 
the limitations of the present approach: 
 
i) We are neglecting the possible back-reaction corrections to 
$\ft$ in (\ref{ft}). This is valid for times below $t_{BR}$, when 
the pion correlations are of the same order as the leading order, 
that is,  $\langle \pi^2 \rangle (t_{BR})\sim f^2$. However, we 
will see in section \ref{fpi} that within our approach one can 
{\it estimate} $t_{BR}$ by calculating the loop corrections to 
$\fpi (t)$. Fortunately, as we said before, nearly  all the relevant 
nonequilibrium pion production physics happens before that time. 
This limitation comes from the fact that we are treating $\ft$ as 
{\it external}, similarly to quantum field theory in an external 
curved space-time, as we have seen in section \ref{rencur}.  A 
self-consistent approach, analogous to treat also the metric 
quantum-mechanically  in a perturbative low-energy fashion would 
be very interesting but is out of the scope of this work. 
 
ii) We are assuming that the system is the late stage of the 
expansion, so that it makes sense to treat the amplitude of the 
oscillations $q$ as a small parameter (narrow resonance 
approximation). That means we will only retain the leading order 
in $q$ and thus ignore $\Od (q^2)$ corrections. This is consistent 
with ChPT to one-loop if $qM^2=\Od(p^2)$. We will see that this 
simplification amounts, among other things, to consider the 
Mathieu equation for the pion modes. 
 
iii) We shall restrict to one loop in ChPT, in the chiral limit 
and for the $SU(2)$ chiral symmetry. As commented before, the 
nature of the ChPT approach  allows to extend our calculations 
including quark masses and three flavours.

\subsection{Dimensional Regularization of the LO propagator.} 
\dla{sec:drlo}

 According to the previous discussion, 
 the differential equation (\ref{homo})  becomes to leading order 
in $q$, the Mathieu equation: 
\be 
\frac{d^2 h_1(z,k)}{d z^2} + (a(k)-2 q \cos 2z) h_1(z,k) =0 
\dla{mathieu} \ee where $z=Mt/2-\pi/4$, $a(k)=4k^2/M^2$ and, 
without loss of generality, we will take $q>0$. 
 
The solutions of   the Mathieu equation are known and tabulated. 
We have collected in Appendix \ref{solmat} some useful results 
about this equation. For our purposes, the most relevant feature 
 is that it admits unstable solutions 
exponentially growing in time for certain values of the parameter 
$a$. This is the simplest version of the parametric resonance 
mechanism. 
 The instabilities arise in bands in $k$, centered at 
 $k_n=nM/2$,  of width $\Delta k_n=\Od(q^n)$ (see Appendix \ref{solmat}). 
 Therefore, in the 
 narrow resonance approximation  we will just neglect the 
 width of all the bands but the first one. 
 A typical unstable solution for the equal-time correlation 
function $G_0(t,t,k)$ has been plotted in Figure 
 \ref{G0} around the first band 
  for a particular choice of the parameters. The 
solutions typically oscillate with an exponentially growing 
amplitude inside the unstable region. These will be the field 
configurations responsible for explosive pion production. 
\begin{figure} 
\centerline{\epsfig{file=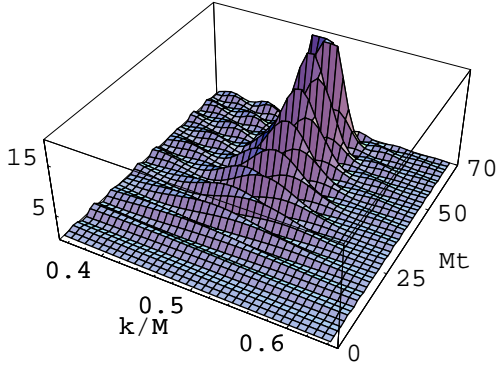,width=11cm}} 
\caption{Profile of  $iMG_0(t,t,k)$ for $T_i=M/100$ 
and $q=0.1$. The instability band for this case lies roughly 
between $0.4M<k<0.6M$.} \dla{G0} 
\end{figure} 
 
 Our next step will be to analyse the equal-time correlator $G_0 
(t)$ in parametric resonance, separating its UV divergent part in 
DR. Therefore, we need to know the behaviour of $G_0 (t,t,k)$ in 
(\ref{g0tk}) for large $k$. Since we are considering only one 
resonant band in momentum space  then $G_0 (t,t,k)$ is in the 
stable region  for large enough $k$. For small $q$, the analytic 
solution  in the stable zone is given in 
 (\ref{cestab}). Let us define 
$k_0$ such that for $k>k_0 (>M/2)$ one can  simply take $\tilde q 
= q/[2(a-1)]$ in (\ref{cesqtil}). As explained in Appendix 
\ref{solmat}, it is not difficult to estimate numerically the 
value of  $k_0$ 
 for a given $q$. We have found that 
$k_0=M$ satisfies the above requirement  within our approximation 
range for $q\leq 1$. Nevertheless, we have checked that our 
results do not depend on the choice of $k_0$ as long as $k_0\geq 
M$. 
 
Thus, replacing the approximate solution (\ref{cestab}) 
in (\ref{f1cese}), solving for the coefficients $A(k)$ and $B(k)$ 
 and replacing the result in(\ref{g0tk}) we find, to leading 
order in $q$, 
 
\ba 
 i G_0 (t,t,k)&=&\frac{1}{2k} \left[1+2 n_B (k) \right] \left\{ 1 
\right.\nonumber\\ 
 &+&\left.\frac{q 
 M^2}{4} \frac{1}{k^2-M^2/4} \left[\sin Mt - 
\frac{M}{2k}\sin 2k t\right]\right\}\nonumber\\ 
 \dla{gotkuv} 
\ea 
for $k\geq k_0$. 
First, let us split the $k$ integral in (\ref{g0def}) as \be 
\int_{0}^\infty = \int_0^{k_0} + \int_{k_0}^\infty \dla{split}\ee 
 
The first piece is finite so that we can take $d=4 $ and use 
either the asymptotic or numerical solutions for the Mathieu 
equation. In the second, which is UV divergent, we replace the 
solution (\ref{gotkuv}). It is clear that the piece proportional 
to the Bose-Einstein function is UV finite, since $n_B(k)$ 
decreases exponentially for large $k$. In addition, the integral 
of any power of $k$ is identically zero in DR, so that 
$\int_{k_0}^{\infty}dk k^{d-3}=-\int_{0}^{k_0}dk k^{d-3}$  and 
 we can absorb that contribution into the first piece in 
(\ref{split}) \footnote{The above result can be understood formally in DR 
by taking $\lim_{m\rightarrow 0}\int_{k_0}^\infty 
k^{d-1}/(k^2+m^2)=-k_0^{d-2}/(d-2)$.}. Therefore, let us write for 
$t>0$:

\ba i G_0 (t)=\frac{T_i^2}{12} -\frac{qM^2}{32\pi^2} I(t) + 
i G_0^{div} (t)\nonumber\\ 
+ \frac{1}{2\pi^2} \left\{ 
\int_0^{k_0} dk k^2 \left[iG_0 (t,t,k)- 
\frac{1}{2k}\left[1+2 n_B (k)\right]\right] 
\right.\nonumber\\ + \left.\frac{qM^2}{4} \int_{k_0}^\infty dk k 
n_B(k) \frac{\sin Mt - \frac{M}{2k}\sin 2kt}{k^2-M^2/4}\right\} 
 \dla{g0fin} 
  \ea 
where we have separated explicitly the equilibrium contribution 
which is just 
\be i G_0^{eq} =\frac{1}{2\pi^2} \int_0^\infty dk k 
\frac{1}{e^{\beta_i k}-1}=\frac{T_i^2}{12}\dla{g0eq}\ee 
and the integral 
 
\be 
I(t)=M\int_{k_0}^\infty dk \frac{\sin 2kt}{k^2-M^2/4} 
\dla{itildet} \ee is finite for $t\geq 0$. It is important to 
point out that if we had chosen a different  phase $\varphi$ in 
$\ft$, as for instance $\ft=f[1-(q/2)(\cos Mt-1)]$, we would have 
found a singular behaviour near $t=0$. In fact, instead of $I(t)$ 
above, we would have an integral which is finite for {\it $t>0$} 
and $d\rightarrow 4$ but logarithmic divergent in the 
$t\rightarrow 0^+$ limit. In this sense, $t$ acts as a natural 
regulator. This would not have been a  limitation to our 
approach, since we are meant to observe the system for times such 
that the $t=0$ effects are unimportant. In fact, we have checked 
numerically that the influence of those terms is irrelevant for 
$Mt\gtrsim 1$. The behaviour at $t=0$ is just a consequence of 
our  non-analytic approach, where the nonequilibrium effects 
appear instantaneously and is a well-known problem in 
nonequilibrium field theory. In fact,  in \cite{baa98} it has been 
pointed out that it can be cured by a suitable choice of the 
initial state, which  for  a time-dependent mass term amounts to 
take $m^2(0^+)$ equal to the initial mass. This is exactly what we 
have done with our choice of phase in (\ref{ft}), since 
$m^2(0^+)=0$ in the chiral limit. Our results confirm the analysis 
in \cite{baa98} from a completely different viewpoint, namely 
working in path integral within the ChPT framework in the DR 
scheme.

The divergent part in (\ref{g0fin}) is given by 
 
\ba i G_0^{div} (t)=\frac{\sin Mt}{\Gamma (\frac{d-1}{2}) 
(4\pi)^{\frac{d-1}{2}}} \frac{q 
 M^2}{4} 
 \int_{k_0}^\infty dk \frac{k^{d-3}}{k^2-M^2/4} 
\dla{g0divin} 
 \ea 
 
 
 We will proceed now to regularize this expression. We have \cite{gr}, 
\ba 
\int_{k_0}^\infty dk 
\frac{k^{d-3}}{k^2-M^2/4}&=&\frac{k_0^{d-4}}{4-d} \ {}_2 F_1 
\left[1,2-\frac{d}{2};3-\frac{d}{2};\frac{M^2}{4 k_0^2}\right] 
\nonumber\\ 
&=& 
\frac{(k_0^2-M^2/4)^{\frac{d-4}{2}}}{4-d} + \Od 
(d-4), 
\dla{g0div1} 
\ea 
%
which replaced in (\ref{g0divin}) yields: 
\ba iG_0^{div} 
(t)&=&-\frac{qM^2}{32\pi^2}\left\{2(k_0^2-M^2/4)^{\frac{d-4}{2}} 
\sin{Mt}\left[\frac{1}{d-4}\right.\right.\nonumber\\ 
&-&\left.\left.\frac{1}{2}\left(\log\pi-\gamma+2\right)\right] 
+\Od(d-4)\right\} 
 \dla{g0divfin}\ea

 As we will see in detail in section 
\ref{fpi}, we will be able to renormalize this divergence in the 
low-energy constant $L_{11}$ so that the answer for the 
observables is finite and scale independent. This means that the 
regularization of the above UV divergence is consistent. 

\section {The pion decay functions $\fpit$ to one loop} 
\dla{fpi}

The pion decay constant acquires one-loop corrections in ChPT. 
Those corrections are finite and scale-independent once the 
contribution from the $O(p^4)$ lagrangian is taken into account 
\cite{gale}. 
 The same will happen in our nonequilibrium model, where the pion 
 decay constant becomes  a time-dependent function $\fpi (t)$, 
 which to tree level is just $\ft$. For us, the importance of the 
  calculation of $\fpi (t)$ is twofold: 
 first, it will provide an explicit check of consistency of our 
 renormalization scheme. 
Second, it will help us to understand the time scales. 
 In particular, 
 the size of the loop correction 
 will define the back-reaction time, when it equals the tree level 
 contribution. For times well below that scale, our approach remains 
 perfectly valid and yields predictions for 
  observables such as $\fpi (t)$ and the pion number. 
  As a matter of fact, the same philosophy is followed at 
 finite temperature in equilibrium, where  in the chiral limit 
 \cite{gale87}: 
\begin{eqnarray} 
\left[\fpi^2 (T)\right]=f^2\left(1-\frac{T^2}{6 \fpi^2}\right) 
\dla{fpieq} 
\end{eqnarray} 
with $\fpi\simeq 93$ MeV. Clearly, this result is valid only for 
temperatures below $T^*=\sqrt{6} \fpi\simeq 228$ MeV. In fact, 
even though $\fpi (T)$ is not a good order parameter for the 
chiral phase transition \cite{boka96}, it should decrease as $T$ 
approaches the critical temperature $T_c$. Therefore, the one-loop 
result  (\ref{fpieq}) already reproduces the correct qualitative 
behaviour and indeed it provides a reasonable estimate $T^*$ for 
the critical temperature. The $T^2$ term in (\ref{fpieq}) is 
nothing but the thermal pion correlator $\langle \pi^2 (0)\rangle$ 
in the chiral limit in (\ref{g0eq}) and at temperatures near $T^*$ 
pion correlations are of the same size as $f^2$ ($\sim \langle 
\sigma^2\rangle$ in the $O(4)$ model) so that higher order 
corrections become equally important. 
 
We should bear in mind that the definition of $\fpi$ is subtle 
even in thermal equilibrium. In fact, it is more convenient to 
define it as the residue of the  axial-axial thermal spectral 
function $<T_C [A_\mu^a (x), A_\mu^b (y)]>$ \cite{boka96,kash94}, 
where $A_\mu^a (x)$ is the axial current ($a=1,2,3$), instead of 
using the PCAC theorem \cite{dogoho92}. In this way, one avoids 
dealing with the reduction formula and asymptotic states at finite 
temperature. In addition, it is important to bear in mind that, 
due to the loss of Lorenz covariance in the 
 thermal bath, one can define two different pion decay constants 
  $\fpi^s$ and  $\fpi^t$ 
 corresponding, respectively, 
  to the   space and time  components of the axial current 
 \cite{pity96}. In fact, the chiral symmetry imposes relations between them 
 and  the in-medium pion dispersion law. If the chiral 
 symmetry is exact (as it is in our case) pions remain massless 
 but their velocity  $v_\pi$ can be less than the speed of light and their 
 thermal width can be different from zero. The relation with 
 $\fpi^{s,t}$ is given by 
 $v_\pi\simeq \re \fpi^s/\re \fpi^t$, while the thermal width is 
 proportional to the imaginary parts of $\fpisp$ and $\fpite$ 
 \cite{pity96}. Nevertheless, to one loop in ChPT one has 
 $\im\fpisp=\im\fpite=0$ and 
 $\fpi^s=\fpi^t=\fpi (T)$ in 
 (\ref{fpieq}) although there are corrections to the velocity beyond one loop 
 \cite{pity96}. As we are going to see, in the present model  we will 
 get a small but nonzero difference $\fpisp-\fpite$ to one-loop, 
 unlike equilibrium, which could be interpreted as 
  a small nonequilibrium deviation  for the pion velocity. 
 The   pion velocity plays also an important role in the hydrodynamics 
 of the 
 chiral phase transition  \cite{son00}.

\subsection{Nonequilibrium pion decay functions} 
\dla{sec:nepddf}  All the above considerations for $\fpi$ can be 
extended to nonequilibrium. We refer to our earlier work 
\cite{agg99} for further details. The axial current from 
(\ref{nlsmne}) reads 
\be 
A_\mu^a\vxt=i\frac{\ftsq}{4}\tr\left[\tau^a\left( 
U^\dagger\partial_\mu U-U\partial_\mu U^\dagger\right)\right] 
\dla{axcurr} \ee

As in equilibrium, there are  two independent $\fpi^s(t)$ and 
$\fpi^t (t)$. Their definitions are given in \cite{agg99} 
consistently with the Ward identities of  chiral symmetry. To 
leading order in ChPT (tree level with the action (\ref{nlsmne})) 
one has to consider only $\Od(\pi)$ contributions when expanding 
the $U$ fields in the axial current (\ref{axcurr}) 
%
which yields, as it should, $\fpi^s(t)=\fpi^t(t)=\ft$ at tree 
level. 
\begin{figure} 
\hspace*{-1cm} 
 \hbox{\epsfig{figure=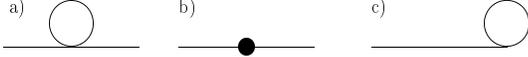,width=9cm}} 
\vspace*{-6cm} \caption {The different diagrams contributing to 
the pion decay functions to NLO in ChPT} \dla{diagrams} 
\end{figure}

To next-to-leading order (NLO) there are three different 
contributions to $\fpi$: the first one comes from the NLO 
corrections to the  propagator. Such corrections are of two types: 
one loop diagrams from the $S_2$ action and tree level ones from 
$S_4$ (diagrams a) and b) in Figure \ref{diagrams} respectively). 
It is important to bear in mind that in the calculation of diagram 
a) one has to integrate over all the branches of the contour $C$ 
in Figure \ref{contour} and the result is independent of $t_i$. 
Second, there is another one-loop diagram involving the product of 
three pion fields at the same space-time point (diagram c) in 
Figure \ref{diagrams}) from the next order in the expansion of the 
axial current. 
Finally, at the same order we have to take into account that the 
axial current in (\ref{axcurr}) is itself modified by the fourth 
order lagrangian in (\ref{lag4gen}) as $[1+f_1(t)] A_0$ and 
$[1+f_2(t)] A_j$ with $f_1 (t)$ and $f_2 (t)$ in (\ref{f12}). 
 
 Once the different contributions have been taken into account, the final 
 result  to $\Od(p^4)$ reads \cite{agg99} 
\ba \left[\fpisp (t)\right]^2&=&f^2 
(t)\left[1+2f_2(t)-f_1(t)\right]-2i G_0(t) \dla{fpispnlo}\\ 
\left[\fpite (t)\right]^2&=&f^2 (t)\left[1+f_2(t)\right]-2i G_0(t) 
\dla{fpitenlo} \ea for $t>0$, with  $G_0 (t)$ the equal-time 
correlation function defined in (\ref{g0def}). 
 The results (\ref{fpispnlo})-(\ref{fpitenlo}) reproduce the 
equilibrium result in (\ref{fpieq}) 
 when the time derivatives of 
$\ft$ are switched off and $G_0^{eq}$ is replaced by (\ref{g0eq}).

As in standard ChPT, the loop corrections to $\fpi$ come directly 
from the equal time pion correlator, which is time-dependent now. 
In addition, there are $\Od (p^4)$ tree level corrections given by 
the terms proportional to $f_1(t)$ and $f_2(t)$. In the chiral 
limit in equilibrium there are no $\Od (p^4)$ tree level 
corrections  because $G_0^{eq}$  is finite.  However, at 
nonequilibrium $G_0 (t)$ is UV divergent as we have seen 
  in section \ref{parres} and the counterterms proportional to 
  $f_{1,2}(t)$ are precisely those needed to arrive to a finite 
  answer. The above result for $\fpi (t)$ should be such that the total 
answer is finite and scale independent because $\fpi$ is an 
observable. Thus, we should be able to absorb the UV infinities 
and the scale dependence (in DR) in the new low-energy constant 
$L_{11}$, as discussed in section \ref{rencur}. In addition, 
 from (\ref{fpispnlo})-(\ref{fpitenlo}) and (\ref{f12}) we see that the 
 difference $[\fpisp (t)]^2-[\fpite (t)]^2$ remains {\em finite} 
 (it depends only on $L_{12}$) 
  so that the same renormalization is valid for 
  $\fpi^{s}(t)$ and   $\fpi^{t}(t)$, which is another consistency 
  check. In fact, the above result yields $\fpisp (t)\neq \fpite (t)$ to 
 one loop, unlike the equilibrium case. Therefore, the 
  plasma expansion induces modifications in the pion velocity 
  larger than in equilibrium. However, note 
 that we are following the equilibrium arguments given in 
 \cite{pity96} in order to relate a nonzero value for 
 $\fpi^s-\fpi^t$ with the in-medium pion velocity. Thus, our 
 conclusions in this respect must be taken with care. In section 
 \ref{numres} we will come back to this point and give some 
 numerical estimates.

\subsection{$\fpisp (t)$ and $\fpite (t)$ in parametric resonance.} 
 
Let us concentrate now in the parametric resonance approach, with 
$\ft$ given in (\ref{ft}) and where we keep only the leading order 
in the amplitude of the oscillations $q$. The loop contribution is 
given by $G_0(t)$, which we have analysed in detail in section 
\ref{parres}. It will grow exponentially in time due to explosive 
pion production, once the infinities have been suitably 
subtracted, whereas the tree level corrections $f_{1,2} (t)$ 
remain bounded in time. 
 
The first step is  to show how the infinities cancel in the final 
answer for  $\fpi(t)$. For that purpose, we replace in 
(\ref{fpispnlo})-(\ref{fpitenlo}) the functions $f_{1,2}(t)$ 
  in (\ref{f12}) to leading order in  $q$. 
On the other hand, in section  \ref{parres} we have regularized 
the equal-time two-point function  in DR. Its divergent 
 divergent part for $d\rightarrow 4$ is given in (\ref{g0divfin}). 
 According to 
our previous discussion, we should be able to absorb the divergent 
part in $L_{11}$. In fact, collecting the piece proportional to 
$L_{11}$ in both $\fpi^{s,t} (t)$ (remember that they only differ 
in terms proportional to $L_{12}$) plus the divergent contribution 
in (\ref{g0divfin}) and using (\ref{l11ren}) yields 
\ba \left[\fpit^{div} \right]^2  &=&-q M^2 \sin Mt \left[12 
L_{11}^r (\mu) \right.\nonumber\\ 
 &+& \left.\frac{1}{16\pi^2} \left( 1+ 
\log\frac{\mu^2}{4k_0^2-M^2}\right)\right] \ea where we have taken 
the  $d\rightarrow 4$ limit.  The above contribution is finite and 
scale independent (the explicit dependence with $\mu$ is 
compensated by that in $L_{11}^r (\mu)$ as explained before) which 
is a very important consistency check of our approach. Notice that 
it is crucial that the  divergent contribution  in 
(\ref{g0divfin})  has exactly the same time dependence as $\ddot 
f(t)/f (t)$ in (\ref{f12}).

Therefore, collecting the various pieces above, we can write the 
final result for $\fpisp(t)$ and $\fpite(t)$ in parametric 
resonance (to leading order in $q$ and to one loop in ChPT) as 
\ba &&\frac{\fpisp (t)}{\fpi (T_i)} = 1-\frac{q}{2}\sin Mt 
+\frac{qM^2}{\fpi^2}\left\{\frac{I(t)}{32\pi^2} 
\right.\nonumber\\ &+&\left.\sin Mt\left[L_{12}-6 
L_{11}^r(\mu)-\frac{1}{32\pi^2}\left(1+\log\frac{\mu^2}{4k_0^2-M^2} 
\right)\right] 
\right.\nonumber\\ 
 &-&\left. \Delta_{B} (t)\right\}-\frac{\Delta_{unst} 
(t)}{\fpi^2}\dla{fpifinal}\ea \ba \fpite (t)=\fpisp (t) \ 
(L_{12}\rightarrow -L_{12})  \dla{fpifinal2} \ea for $t>0$, with 
$\fpi (T)$ in (\ref{fpieq}), $I(t)$ in (\ref{itildet}) and 
\ba \Delta_{B} (t)= \frac{1}{8\pi^2} \int_{k_0}^\infty dk k n_B(k) 
\frac{\sin Mt - \frac{M}{2k}\sin 2kt}{k^2-M^2/4} \dla{deltab}\ea \ba 
\Delta_{unst} (t)&=&\frac{1}{2\pi^2} \int_0^{k_0} \!\!dk k^2 
\left[iG_0 (t,t,k)-\frac{1}{2k}\left[1+2 n_B 
(k)\right]\right]\nonumber\\ &&\dla{deltaunst}\ea

Note that the only unstable (exponentially growing) contribution 
in (\ref{fpifinal}) 
   is 
  given by $\Delta_{unst} (t)$ in (\ref{deltaunst}). The rest is 
  bounded in time. Therefore, according to our previous discussion, 
  we can estimate the back-reaction time as 
   $\Delta_{unst} (t_{BR})\simeq\fpi^2$. 
 Thus, we have an approximate idea of the 
time scale during which our one-loop approach can be trusted. From 
that time onwards, the back-reaction corrections to $\ft$ coming from the 
coupling to 
   the pion fields  cannot be ignored. 
 
   \subsection{Numerical results} 
\dla{numres} 
 
 Since the result is independent of the scale,  we will 
  fix $\mu=1$ GeV and use the numerical values of $L_{11}$ and 
  $L_{12}$ determined phenomenologically in \cite{dole91} and 
  given in section \ref{rencur}. We have also taken $k_0=M$ (see our 
  previous comments). 
 
There are  still three parameters we have to fix corresponding to 
the initial conditions: the initial amplitude of the oscillations 
$q$, the initial frequency $M$ and the initial temperature $T_i$. 
According to our previous comments, 
   $M$ should be around the value of the 
   $\sigma$ mass in the $O(4)$ model, although  in the present 
 approach it is not 
   necessary to assume the existence of a $\sigma$ particle. We have 
   considered three different cases: 
   $M=$ 0.1, 0.6 and 1 GeV. The second value is 
   the one more often used in the literature and also the closest one 
   to the recent determinations of $m_\sigma$ \cite{olleretal}. 
 As for the initial temperature, we have fixed for 
definiteness $T_i= $ 50 MeV.  The standard approach is to assume 
an initially supercooled (zero temperature) state when the 
$\sigma$ field starts its rolling down from the top of the 
potential. In this process the initial potential energy is 
converted into thermal energy of the pion gas (reheating) 
\cite{mm95}. Hence,  it is reasonable to assume in our case a 
nonzero but small initial temperature. Even though our model does 
not have a direct interpretation in terms of an effective 
potential, it corresponds in the language of the $O(4)$ model to 
an initial condition where the $\sigma$ has already reached the 
bottom of the potential and it is oscillating around it. 
Nonetheless we must point out that the results depend very weakly 
on $T_i$. The reason is that in the right hand side of 
(\ref{fpifinal}) the dependence with $T_i$ enters through the 
integrals (\ref{deltab})-(\ref{deltaunst}) which are dominated by 
contributions near $k\simeq M$ and are therefore strongly damped 
by  the Bose-Einstein distribution $n_B$  for temperatures $T_i\ll 
M$. We have checked that taking different values for $T_i$ in the 
range 10-100 MeV the curves showed below remain almost unchanged. 
Finally, we have considered $q=0.1$ and $q=0.2$ to illustrate the 
dependence of our results with the initial amplitude. The initial 
values we are 
 considering here are similar to those used in the literature 
\cite{hiro,mm95,kaiser}.

   The results for  $\Delta_{unst} (t)/\fpi^2$ 
   are plotted in Figure \ref{unst}. 
   The values of the estimated $t_{BR}$ are also 
    given in that figure.    One  clearly observes that increasing either $q$ 
    or $M$ makes the unstable modes grow faster and 
    overcome earlier the tree level value. In fact, the upper envelope of the 
     long-time 
    oscillations is proportional to $(qM^2/(4\pi\fpi)^2) \exp 
    (q Mt/2)$ since, from 
    (\ref{Funst}) and (\ref{g0tk}), 
    the dominant exponential contribution to the two point function at 
    long times is $\exp(2\mu z)$, the maximum of the Floquet exponent 
    $\mu$ in (\ref{muunst}) being $\mu\simeq q/2$ 
     at the center of the unstable 
    band, which has  width  $q$. 
 
    The time it takes 
    for the pion correlations to overcome the tree 
    level has been estimated  in the $O(4)$ model in this regime 
and it 
lies between  5-10 fm$/c$ \cite{hiro,kaiser}. 
     As  explained before, this is the same time 
    scale as that when dissipation makes pion production stop. 
In the four cases we have considered, it is 
    clear that our approach remains valid for the time relevant to 
    pion production. For definiteness, we will restrict 
 from now on to the choice of parameters 
     (d) in Figure \ref{unst} giving $t_{BR}\simeq$ 10 fm/$c$, which 
     is of the order of the plasma lifetime. In turn, note that 
     for cases (c) and (d) we have $qM^2/\Lambda^2_{\chi} \simeq 0.07$ 
and $qM^2/\Lambda^2_{\chi}\simeq 
     0.05$ respectively, so that our ChPT approach 
to $\Od (q)$  is perfectly valid. 
 
\begin{figure} 
\hspace*{-.5cm}\centerline{\epsfig{file=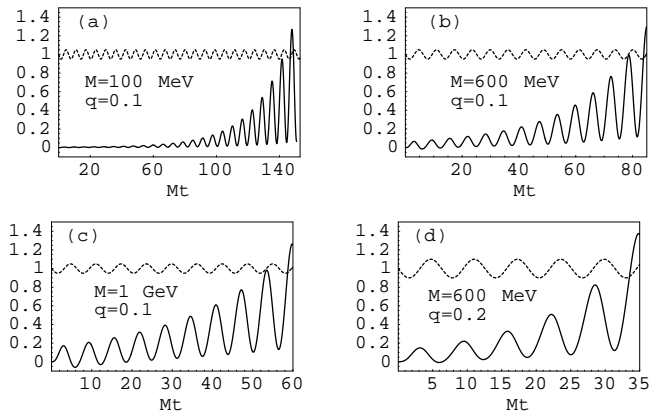,width=11cm}} 
\caption{The solid line is $\Delta_{unst} 
(t)/\fpi^2$. The dashed line is the tree level contribution 
$1-q/2\sin Mt$. $T_i=$ 50 MeV for 
all cases. The values of the back-reaction time are approximately 
given by: (a) $t_{BR}\simeq$  300 fm$/c$, (b) 
$t_{BR}\simeq$ 28 fm$/c$, (c) $t_{BR}\simeq$  11 fm$/c$ and 
 (d) $t_{BR}\simeq$  10 fm$/c$} 
\dla{unst} 
\end{figure}

In Figure \ref{fpifig} we have depicted the total result for 
$\fpisp (t)/\fpi(T_i)$ in (\ref{fpifinal}) for the same 
cases as in Figure \ref{unst}. In this curve, the upper limit in 
time corresponds  roughly to the onset of the back reaction. Its effect is to 
make the amplitude of the oscillations grow, whereas for 
$t<t_{BR}$ the amplitude remains approximately constant. 
As a matter of fact, one could wonder whether  the oscillations of 
the unstable part could cancel those of the tree level in 
(\ref{ft}) so that $\fpi (t)$ remains roughly constant at long 
times, which would be interpreted as  a dissipation effect. 
Clearly, this is {\it not} the case as it can be seen in Figures 
\ref{unst} and \ref{fpifig}. The oscillations coming 
$\Delta_{unst} (t)$  have indeed the same frequency as the 
tree level ones but their phase is different as it can be seen  in 
Figure \ref{unst}. In fact, in that figure one observes that for 
long times the phase of $\Delta_{unst}$ is shifted almost $\pi/2$ 
with respect to the tree level and therefore the contribution with 
$\cos Mt$ dominates. \footnote{Working out the 
  expressions for the unstable band 
given in Appendix \ref{solmat} it can be seen that there are three 
different  terms  proportional to the 
  leading long-time exponentials $\exp(qMt/2)$ in $G_0 (t,t,k)$. The first one is 
   proportional to $\sin Mt$ and  changes sign, 
 to leading order in $q$, under a 
reflection with respect to the band center. The other two are, 
respectively,   time-independent  and 
  proportional to $\cos Mt$ and they are symmetric under such reflection. 
Therefore, by  integration in $k$, 
 the $\sin Mt$ term is 
 suppressed by at least one power of $q$ with respect to the other 
 two. 
 The time-independent term increases 
 the central value of the 
 oscillations, as seen  in Figure \ref{unst}.} 
 In other words, the effect of the pion correlations itself is 
not enough to make the system equilibrate. This is consistent with 
the analysis in \cite{boyaetal96,boyaetal95} where it is 
shown that such dissipation appears only when the back-reaction is 
self-consistently included, which we have not done, as explained 
before. This issue will be confirmed by the 
 analysis in the next section, where it is 
shown that the pion number grows exponentially in time even when 
the loop corrections are included. If the back-reaction was 
considered, the pion number should reach a maximum value and then 
stop growing \cite{boyaetal96}. Our $\ft$ is an external force 
whose shape is not changed during the time evolution, which is 
consistent only below $t_{BR}$. Furthermore, not only the phase of 
the long-time oscillations is different, but, as it can be seen 
from Figure \ref{fpifig}, the central value also decreases with 
time (increases in Figure \ref{unst}). This is due to the constant 
term proportional to $\exp(qMt/2)$ commented before. We interpret 
this effect as a reheating of the system \cite{mm95}. The fact 
that $\fpi$ decreases with temperature gives support to this idea. 
 Also, assuming that when the system reaches the back-reaction 
time scale we can use approximately the equilibrium expressions 
then, according to (\ref{fpieq}), an estimate of the final 
 temperature is given by 
\be 
T_f^2\simeq  6\left(\fpi^2-\overline{\fpisp} (t_f)^2\right) \ee 
where the bar denotes  time average: 
\be 
\overline {F} (t_f)=\frac{1}{t_f}\int_0^{t_f} dt' F (t') 
\dla{tav}\ee 
 
For the above estimate of $T_f$ it is not important whether we choose 
$\fpisp$ or $\fpite$. 
 Thus, for the parameters in case (d) in Figures \ref{unst} and 
\ref{fpifig} we obtain, for $M t_f\simeq 25-30$,  $T_f\simeq  125-140$ 
MeV which is not far from  experimental determinations of the 
freeze-out temperature \cite{ris01}.  
 Recall that this value is almost independent of the initial 
 temperature $T_i$ and it is therefore compatible with a 
 supercooled initial state. Although our estimate is based on 
 assumptions about the final state, we are using 
  $\fpi$ and not the  pion distribution function. The reason is that 
 $\fpi$ always remains close to its equilibrium value, according to 
 our previous discussion, unlike the distribution function which 
 is not thermal, as it will be shown in the next section.

\begin{figure} 
\centerline{\epsfig{file=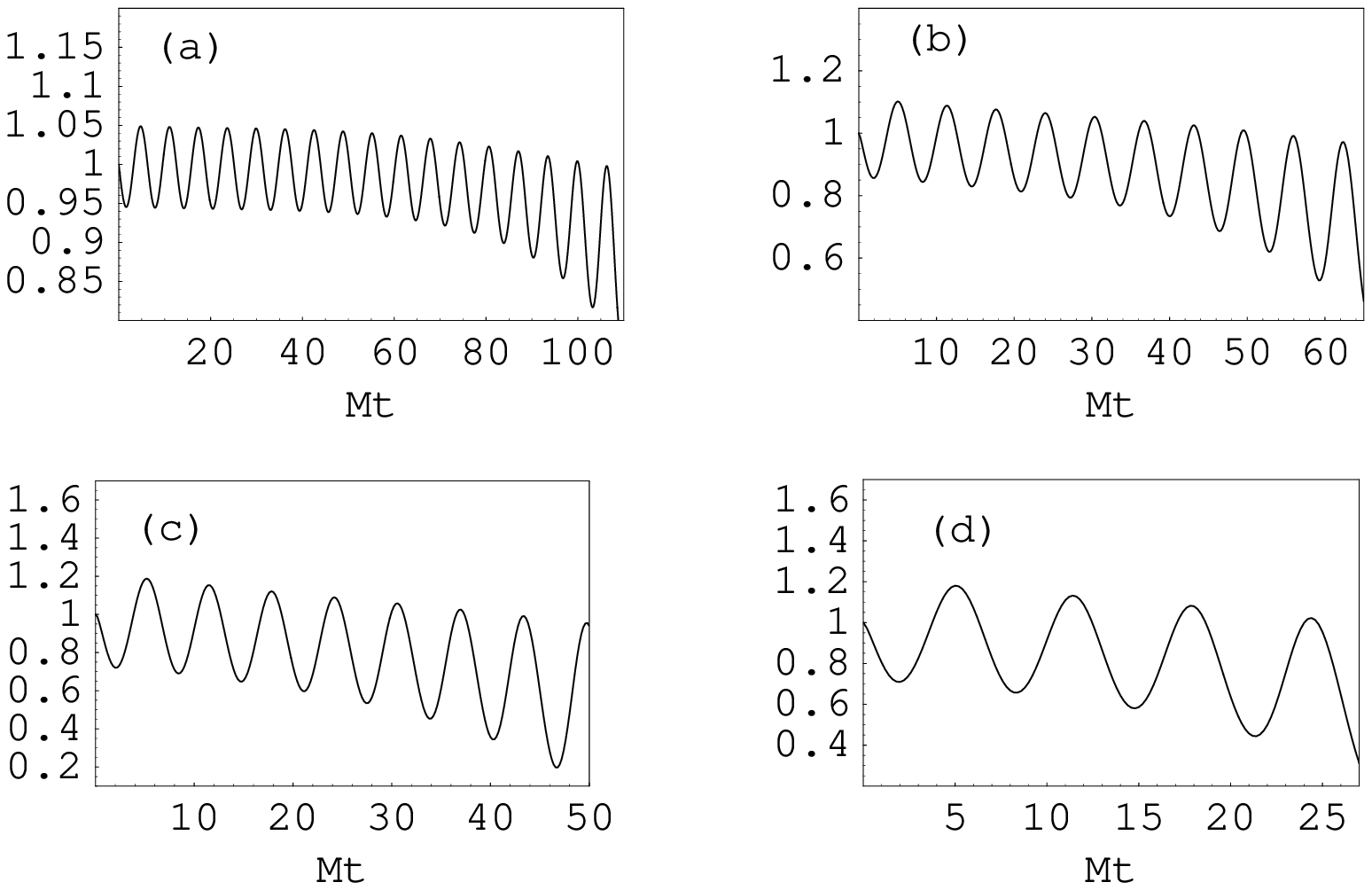,width=11cm}} 
\caption{$\displaystyle\frac{\fpisp (t)}{\fpi(T_i)}$ for the same 
cases as in figure \ref{unst}} \dla{fpifig} 
\end{figure}

Next we will come back to the issue of the pion velocity. To the 
order we are considering, we have, from (\ref{fpifinal}) and 
(\ref{fpifinal2}): 
\be 
\frac{\fpisp (t)}{\fpite (t)} \simeq 1+2 L_{12}\frac{qM^2}{\fpi^2} 
\sin Mt 
\dla{timedepvel} 
\ee 
 
It is unclear whether one can extrapolate from the analysis in 
\cite{pity96} and thus identify the above with the pion velocity 
$v_\pi$. In fact, note that $L_{12}<0$ and then $v_\pi>1$ whenever 
$\sin Mt<0$. This is a similar problem as trying to identify 
$m^2(t)$ with a pion mass which, as we have emphasized,  is not 
correct 
 since the chiral symmetry is exactly preserved and the pions 
  remain massless. The problem, as it is discussed below, is 
trying to define a time-dependent dispersion law, which is 
meaningless unless additional restrictions are imposed, such as 
adiabaticity. However, and following our previous argument, by the 
final time where the equilibrium expressions are meant to be 
approximately valid, an estimate of the maximum variation of 
$v_\pi$ would be given by taking the time average of the above 
quantity. Thus we get ${\Delta v_\pi}^{\!\!\!max}\simeq 0.003$ for 
$M t_f=30$, for  the same  parameters as before. Note that this 
correction is even smaller than what is expected from ChPT beyond 
one loop, namely $\Delta v_\pi\simeq$ 0.13 for $T=$ 150 MeV 
\cite{pity96}.

\section{The pion number} 
\dla{number}

\subsection{The nonequilibrium particle number. Definitions} 
\dla{sec:parnum} 
 
One should bear in mind that the concept of particle number out of 
equilibrium is rather subtle. The nature of the problem is well 
illustrated once more in 
  curved space-time QFT \cite{bida82}. The particle number depends on the 
reference frame and thus 
 the initial vacuum state may contain 
particles during its subsequent time evolution. In other words, 
the state which is regarded as the vacuum at time $t$ is different 
from that at $t=0$. 
However, the particle number can be given a physical meaning in 
some particular cases. For instance, if the metric is  conformally 
 Minkowskian (as the RW spatially flat metric we are considering here) or it 
  is Minkowskian on the space-time boundary. 
   Another interesting regime is when the expansion rate is 
  small compared to the typical frequencies involved, which is the 
  so-called adiabatic limit. 
  Similarly, in nonequilibrium field theory one has to specify the 
state with respect to which particles are defined. 
 One possibility is to choose the initial Fock space, which is 
 simpler 
 if the initial state is the equilibrium one (the analog of the Minkowski 
  limit). In that case, one considers 
 the time evolution of the initial creation and annihilation operators with 
 the time-dependent density matrix \cite{boyaetal96,boyaetal95}. Physically, 
 this corresponds to the number of {\it initial} particles. In our 
 case these are massless pions. Since our model preserves the 
 chiral symmetry at all times, pions remain massless for $t>0$. 
 Given the difficulties related to the definition of the  nonequilibrium 
 dispersion law (see comments below),  we 
 will restrict here to the number of massless pions. 
  Another possibility, when dealing with time-dependent mass terms 
like the one in (\ref{act2pi}) is to define the adiabatic number, 
where the dispersion law is assumed to be simply 
$\omega_k^2(t)=k^2+m^2(t)$.   The particle number is then defined 
in terms of the  creation and annihilation operators that 
diagonalize the instantaneous Hamiltonian via a Bogoliubov 
transformation \cite{cooper9596}. \setcounter{footnote}{0} We 
remark that the adiabatic limit is consistent for a slowly varying 
$m^2(t)$ and in fact it can be defined only for real $\omega_k 
(t)$. In our case this 
approximation would be valid only for $k^2>qM^2/2$. For small $q$ 
this would capture anyway the resonance band and hence the 
parametric amplification. However, as we will see below, it is not 
even clear whether  the dispersion law can be assumed to be 
adiabatic for all times. Both definitions of particle number 
coincide at $t=0$.

The nature of our approach makes it more suitable to define the 
particle number  in terms of correlation functions, rather than in 
the canonical formalism. To illustrate the way we will proceed, 
let us consider first a free scalar field $\phi$ of mass $m$ in 
thermal equilibrium, so that $n(k)$  is just given by the 
Bose-Einstein distribution function (\ref{be}). The 
time-independent Hamiltonian of the system is \be H=\int 
d^3\vec{x} \frac{1}{2} \left[ {\dot \phi}^2 + (\nabla 
\phi)^2+m^2\phi^2\right]\dla{freeham}\ee 
 
The  thermal averaged  energy of the system 
per unit volume  is related to the 
particle number $n(k)$ simply as 
 
$$ \frac{\ll H\gg}{V}=\int\frac{d^3\vec{k}}{(2\pi)^3} \omega (k) 
\left[n(k)+\frac{1}{2}\right] $$ where 
 $k\equiv\vert \vec{k} \vert$ and  $\omega^2 (k)=k^2+m^2$ is the free 
 dispersion law. Thus, from 
 (\ref{freeham}) and  the definitions of the 
two-point functions in section \ref{lop}, 
\ba n(k)+\frac{1}{2}&=&\frac{i}{2\omega (k)} \left[ 
\left.\frac{d}{dt_1}\frac{d}{dt_2}G_0^> 
(t_1,t_2,k)\right\vert_{t_1=t_2}\right.\nonumber\\ 
 &+&\left. \omega^2 (k) 
G_0^> (t_1=t_2,k)\right]\nonumber\\ &=&\frac{i}{2\omega (k)} 
\left[-\ddot G_0^> (0,k)+ \omega^2 (k) G_0^>(0,k) \right] 
\nonumber \ea  where $G_0$ is the free propagator,  the dot 
denotes time derivative and we have used time translation 
invariance (thermal equilibrium) so that $G_0^> (t_1,t_2,k)=G_0^> 
(t_1-t_2,k)$,  meaning that the particle number is 
time-independent in equilibrium. Note also that $G_0^>$ above 
actually stands for 
$G_0^{>11}(x,x')=\langle\phi(x)\phi(x')\rangle$ since we are 
taking all time arguments in the $C_1$ branch in Figure 
\ref{contour}. Using $\left[\partial_t^2+k^2+m^2\right] 
G_0^>(t,k)=0$ (the equation of motion) 
 we have: $$ 
n(k)=i\omega(k)  G_0^>(0,k)-\frac{1}{2}=\frac{1}{e^{\beta\omega 
(k)}-1} $$ where we have  used the solution for the free 
propagator in equilibrium which can be read off from (\ref{losol}) 
with $h_1 (t,k)$ in (\ref{twoeqsol}), i.e, \ba i 
G_0^{>eq}(t-t',k)&=&\frac{1}{2k}\left\{\left[1+n_B(k)\right]e^{-ik(t-t')}\right.\nonumber\\ 
&+&\left.n_B(k) e^{ik(t-t')}\right\} \dla{g0+eq} \ea 
 
Our next step will be to extend the above definitions to the 
nonequilibrium case. Thus, following the same steps, we will 
define the nonequilibrium particle number $n(k,t)$ through: 
\be 
 \frac{\langle E(t)\rangle}{V}=N\int\frac{d^3\vec{k}}{(2\pi)^3} \omega (k,t) 
\left[n(k,t)+\frac{1}{2}\right] \dla{nepndef}\ee where $\omega 
(k,t)$ is (formally) the  dispersion law, $\langle E(t)\rangle$ is 
the total energy of the system and $N$ is the number of particle 
flavours. In our case $N=3$, the number of different pions. In 
fact, we should consider a more general definition involving a sum 
over all internal indices but we will see below that to the order 
we are considering, all our expressions remain diagonal in isospin 
space. Thus,  the pion density for a given pion type is 
\be 
\frac{\langle n(t)\rangle}{V}= \int \frac{d^{d-1} k}{(2\pi)^{d-1}} 
n(k,t) \dla{totpndef} \ee where we have kept the space-time 
dimension $d$ in order to regulate the UV behaviour (see below). 
Several remarks are in order here:  As explained before, both the 
particle number and dispersion law may depend on time. In 
particular, loops can introduce  corrections to the tree level 
dispersion law. Furthermore, whenever the time dependence appears 
through the interaction with an {\it external} source (as it is 
the case here) the energy $E(t)$  is not conserved (it is time 
dependent) as it happens  in curved space-time when the 
back-reaction effects on the metric are ignored. Note that $E(t)$ 
is the contribution to the energy from the pions only and the 
oscillations of $\ft$ transfer energy to the pions making the pion 
number $n(t)$ grow with time. 
 
In fact, and following once more the curved space-time analogy we 
have previously discussed, we will calculate the expectation value 
of the total energy through: 
\be 
\langle E(t)\rangle =\int d^3\vec{x} \sqrt{-g} \langle T_0^0 
(\vec{x},t) \rangle \dla{endef} \ee where  $T_{\mu\nu}$ is the 
energy-momentum tensor defined in terms of the lagrangian as 
\cite{weinberg} 
\be 
T_{\mu\nu}=\frac{2}{\sqrt{-g}}\frac{\delta\left(\sqrt{-g} {\cal 
L}\right)}{\delta g^{\mu\nu}}. \dla{tmunudef} \ee 
 
In Appendix \ref{ap:curved} we have reviewed some useful results 
regarding the calculation of the classical energy-momentum tensor 
for the case of interest here (spatially flat RW metric). We 
remark that, by construction, $T_{\mu\nu}$ is symmetric and 
classically conserved, i.e, $T^\nu_{\mu;\nu}=0$  where $;$ denotes 
the covariant derivative. However, this does {\it not} imply 
necessarily that the energy (\ref{endef}) is time-independent. The 
reason is that $P_\mu=\int d^3\vec{x} \sqrt{-g} T_{\mu}^{\ 0}$ is 
not a covariant four-vector \cite{weinberg}. Note also that in 
(\ref{endef}) we are assuming that the expectation value of 
$T_{\mu\nu}$ has perfect fluid form, which will be the case here 
(see below). 
 
 Therefore, we will proceed by computing the energy-momentum tensor to a given 
order and then calculate the number of particles through 
(\ref{nepndef}) and (\ref{endef}). This means to deal with 
expectation values of products of fields at the same space-time 
point and therefore divergent. We will follow the approach of 
point-splitting the fields so that the results are written in 
terms of Green functions. It is important to stress that, as it 
will be seen below, as long as we use dimensional regularization 
the final expressions for the particle number are automatically 
finite, without any need for extra renormalizations of the 
energy-momentum tensor \cite{bida82,cacoja70,lava87}.

 The simplest example of such point-splitting is provided by 
  our previous derivation of the equilibrium particle number 
for a scalar theory, where we have replaced for instance: 
\ba \ll {\dot \phi}^2 \vxt \gg=\lim_{(\vec{x}',t')\rightarrow 
(\vec{x},t) } \frac{d}{dt'}\frac{d}{dt} 
iG^>(\vec{x}-\vec{x'},t-t') \dla{pseq} 
 \ea 
 
We will proceed in the same way for the nonequilibrium case. 
However, as we will see in the next section, sometimes we will 
have to deal with a field structure in the classical $T_{00}$ 
which is not symmetric under field exchange, like ${\dot 
\phi}\phi$. In these cases we will symmetrize first the classical 
expression and then point-split the fields \cite{bida82}, i.e, \ba 
\lefteqn{\langle A_1(x) A_2(x) \dots A_n(x)\rangle} &&\nonumber\\ 
 &=&\frac{1}{n!} 
\lim_{x_j\rightarrow x} \left[\langle A_1(x_1) A_2(x_2)\dots 
A_n(x_n)\rangle \right.\nonumber\\ &+&\left. \langle A_2(x_2) 
A_1(x_1)\dots\rangle+\dots \right] \dla{pssym} \ea where the $x_j$ 
are space-time points, the $A_j$ is a shorthand notation to denote 
either the field or an arbitrary number of its derivatives and the 
dots denote all possible permutations. Note that we are dealing 
 only with boson fields, which are symmetric under field exchange at 
different points. As it will become clear below, symmetrizing the 
fields in this way yields consistent results for the particle 
number.

Another problem  in connection with the point-splitting is the 
$T_C$-ordering of the fields. For instance, we will find 
four-field contributions to the energy-momentum tensor and  we 
need to relate them with $T_C$-ordered four-point Green functions, 
so that we can use Wick's theorem to write the result only in 
terms of two-point functions. We did not have this problem in our 
previous derivation of the equilibrium particle number, since we 
were only dealing with products of two fields. Thus, we need to 
specify how the time arguments of the fields are ordered when 
taking the $t_j\rightarrow t$ limit.  We will use the 
 following prescription \cite{cacoja70,lava87} 
 
\begin{itemize} 
 
\item First we will symmetrize 
 over all possible ways of ordering the 
  classical fields, as it is  shown in (\ref{pssym}). 
 
\item Next, for a given ordering of the classical fields with an 
arbitrary number of field derivatives, we will replace: 
\ba \lefteqn{\langle \partial^x_{\mu_1}\phi (x) 
\partial^x_{\mu_2}\phi (x) \phi (x) \dots \rangle} \nonumber\\ 
&=&\lim_{x_j\rightarrow x}^{\ \ \ \ *} 
\partial^{x_1}_{\mu_1}\partial^{x_2}_{\mu_2}\dots \langle 
\phi (x_1) \phi(x_2) \phi(x_3)\dots\rangle\nonumber\\ &=& 
\lim_{x_j\rightarrow x}^{\ \ \ \ *} 
\partial^{x_1}_{\mu_1}\partial^{x_2}_{\mu_2}\dots \langle 
T \phi (x_1) \phi(x_2) \phi(x_3)\dots\rangle\dla{psT} \ea 
 where 
$\lim^*_{x_j\rightarrow x}$ means to take the $x_j\rightarrow x$ 
limit keeping the time arguments ordered from left to right, i.e, 
$t_1>t_2>\dots$ (all $t_j$ are in the $C_1$ branch of the contour 
in Figure \ref{contour} so that the contour ordering $T_C$ becomes 
the ordinary time-ordering $T$). Note that the field derivatives, 
when present, are pulled out of the $T$-ordering 
\cite{cacoja70,lava87}. 
\end{itemize}

Note that  in the second step we are   choosing a particular way 
to take the limit. If such limit exists, the answer should be the 
same regardless of the order. In this respect it is important to 
bear in mind that expectation values of products of fields (like 
$G^>(x,x')$ and its derivatives) have always a well-defined 
equal-time limit, unlike $T$-ordered products where one has to be 
careful with ill-defined expressions such as  $\delta (0)$, 
$\delta' (0)$ and so on, when taking time derivatives. For 
instance, suppose that we wanted to use the above prescription 
with $\langle\dot\phi\phi\rangle$ to relate it with 
$G(x_1,x_2)=\langle T\phi(x_1)\phi(x_2)\rangle$. Then, \ba 
\lefteqn{\langle\dot\phi(t)\phi(t)\rangle =\frac{1}{2} 
\langle\dot\phi(t)\phi(t)+\phi(t)\dot\phi(t)\rangle} &&\nonumber\\ 
&=&\frac{1}{2}\lim^*_{t'\rightarrow 
t}\left(\partial_t+\partial_t'\right) 
G^>(t,t')=\frac{1}{2}\lim^*_{t'\rightarrow 
t}\left(\partial_t+\partial_t'\right) G(t,t')\nonumber\\ 
 \ea 
where for simplicity we have omitted spatial arguments, which do 
not play any role here, and $\lim^*$ means taking the limit so 
that $t>t'$ and then the last step in the above equation holds. 
 Had we taken the limit keeping $t'>t$, the answer would have been 
the same by continuity of $G^>(t,t')$ and its derivatives. In 
fact, note that this is equivalent to take  $t$ and $t'$ in $C_2$ 
in Figure \ref{contour} which gives the same answer for $G^>$ as 
taking them in  $C_1$ in the $t'\rightarrow t$ limit 
\cite{sewe85}. Of course,  we could have written the result 
directly in terms of $G^>$ without specifying the order of the 
time arguments and the answer is the same either way. In other 
words, Wick's theorem is trivial for two-point functions. However, 
for four-point functions the above continuity arguments apply as 
well and hence this prescription will allow us to use Wick's 
theorem in the standard way for $T_C$-ordered products 
\cite{lebellac}.

\subsection{The  number of pions in parametric resonance} 
\dla{sec:pionnumpr} 
 
\subsubsection{Leading order} 
\dla{sec:numberlo} 
 
Let us start with the lowest order in ChPT. As explained before, 
to lowest order it is enough to consider the lagrangian in 
(\ref{Scurv}) with the RW metric. Furthermore to leading order we 
only need the $\Od (\tilde\pi^2)$ terms in that lagrangian. That is, the 
"free" lagrangian given in (\ref{act2pi}) for the $\pi (x,t)$ 
fields. The reason why we can neglect $\Od (\tilde\pi^4)$ terms to 
leading order when calculating $\langle E \rangle$ can be 
understood in terms of Feynman diagrams. What we are doing is 
starting from diagrams with a given number of vertices of 
different types and closing them in all possible ways. For 
instance, at tree level there is only one point and hence one 
vertex. Thus, the contribution from the $\Od(\tilde\pi^{2n})$ is  a 
$n$-loop closed diagram and hence according to Weinberg power 
counting theorem \cite{we79} it contributes as $\Od(p^{2n})$ in 
the chiral power counting. 
 
Therefore, using the form (\ref{Scurv}) for the $S_2$ action in 
the parametrization of the $\tilde\pi^a\vxt$ fields, we have: 
 
$$ S_2[\tilde\pi]= \intcc \frac{1}{2} \sqrt{-g} 
g^{\mu\nu}\partial_\mu \tilde\pi^a \partial_\nu \tilde\pi^a + 
\Od(\tilde\pi^4)$$ 
 
Hence, the energy-momentum tensor (\ref{tmunudef}) to lowest order 
 reads simply 
 
$$ 
T^{(2,2)}_{\mu\nu} 
=\partial_\mu 
\tilde\pi^a\partial_\nu\tilde\pi^a  -\frac{1}{2}g_{\mu\nu} 
\partial^\alpha \tilde\pi^a\partial_\alpha\tilde\pi^a$$ 
where we have  used  (\ref{vardet}) and the superscript 
$(n,m)$ in $T_{\mu\nu}$ 
 means a contribution coming from the 
$S_n$ action with $m$ pion fields. The above result is the 
standard kinetic term in curved space-time. Using the equations of 
motion to this order 
$\left(g^{\mu\nu}\partial_\nu\tilde\pi^a\right)_{;\mu}=0$ it is 
straightforward to check that $T^{(2,2)}_{\mu\nu}$ is covariantly 
conserved. Besides, from the particular form of 
$T^{(2,2)}_{\mu\nu}$ above it is not difficult to see that its 
expectation value has perfect fluid (diagonal) form, that is, 
$\langle T_{ij} \rangle=0$ for $i\neq j$ and $\langle T_{i0} 
\rangle=0$, just from spatial translation invariance. 
 
Now, let us consider the above for the RW metric. The total 
energy defined in (\ref{endef}) reads to this order 
\ba \langle E^{(2,2)}(t)\rangle =a^2(t) \int \! d^3\vec{x} \ 
\frac{1}{2} \left\langle \left[\dot{\tilde\pi^a}\vxt\right]^2 
+\left[\nabla\tilde\pi^a\vxt\right]^2\right\rangle \nonumber \\= 
\frac{1}{2} \int d^3\vec{x}   \left\langle 
\left[\dot{\pi^a}\right]^2+\left[\nabla\pi^a\right]^2+\frac{\dot 
a^2(t)}{a^2(t)}\left[\pi^a\right]^2 -2\frac{\dot a 
(t)}{a(t)}\pi^a\dot\pi^a\right\rangle\nonumber\\\dla{e22}\ea where 
in the last line we have written the result for the $\pi^a\vxt$ 
fields. The first line above simply states that the energy is 
conformally equivalent to the Minkowski (equilibrium) result in 
the $\tilde\pi$ parametrization. Recall that the spatially flat RW 
metric is related to the Minkowski one by a conformal 
transformation. 
 
Note also that the energy density is {\it not} obtained just by 
replacing $m^2\rightarrow m^2(t)$ for a free scalar theory, as one 
could have expected from the lagrangian (\ref{act2pi}). That would be 
equivalent to work in the adiabatic limit and it would have been the 
answer  defining the energy as  $H(t)=\int d^3\vec{x} {\cal H}\vxt$ 
 with the Hamiltonian density 
${\cal H}=\dot\pi(\partial {\cal L}/\partial \dot\pi)-{\cal L}$. 
However, these two definitions are 
not equivalent in the presence of the external force 
$a(t)$. In fact it can be checked that 
adding to the lagrangian a total derivative 
$\sqrt{-g}{\cal 
L}\rightarrow \sqrt{-g} \left({\cal L}+A^\mu_{;\mu}\right)$ with 
$A^\mu$ a contravariant vector (the action has to remain a scalar) 
which is a functional of the field and its derivatives, does not 
change neither the equations of motion nor the energy-momentum tensor 
defined as (\ref{tmunudef}). However, ${\cal H}$ may change under such 
transformation. For instance, considering the expression for 
the action  $S_2[\pi]$ before  integrating by parts 
 to get  (\ref{act2pi}), i.e,   $S_2[\pi]=(1/2)\int 
 \left( \partial_\mu \pi \partial^\mu \pi +(\dot a/a)^2 \pi^2-2(\dot 
   a/a)\dot\pi\pi\right)$, one gets   ${\cal H}=(1/2)\left( 
   \dot\pi^2 + (\nabla \pi)^2-(\dot a/a)^2 
   \pi^2\right)$. Therefore in that case the ``mass'' term would be 
 proportional to $-(\dot a/a)^2$ rather than  $-\ddot 
 a/a$. In this case $A^0=(\dot a/a)\pi^2$. 
As we have said before, none of these expressions for 
 the energy density is time-independent. 
 
Nevertheless, there is a way to check the 
 consistency of the above result which in fact gives us a hint of how 
 to include the back-reaction effects. If we let $a(t)$ (or $\ft$) 
be a classical 
 field (independent of $\vec{x}$ for simplicity) in the lagrangian, 
 then, by considering also its equation of motion, it can be checked  that 
 $\langle \dot E^{(2,2)}(t)\rangle=0$. Moreover, $H(t)$ 
 coincides with (\ref{e22}) when the extra term coming from ${\cal 
   H}\rightarrow {\cal H} +\dot a (\partial {\cal L})(\partial \dot 
 a)$ is considered. 
From this point of view, in the present approach we are just 
considering that the  field $a$ does not receive quantum corrections 
and its equation of motion is dominated by the kinetic term, which 
 depends only on $g_{\mu\nu}$ but not on the pion fields. That 
term is nothing but the counterpart of the Einstein-Hilbert action, 
 yielding the Einstein equation for $g_{\mu\nu}$ when no matter 
fields are present \cite{weinberg}.

At this point let us recall that we are considering only the 
leading order in $q$ consistently with our power counting. 
Therefore we can neglect the $\dot a^2$ term in (\ref{e22}). Now, 
let us apply our point-splitting prescription to the remaining terms 
 in (\ref{e22}). We will write the result in terms 
of  $h_1(t)$ in 
 (\ref{losol}). Let us consider first the term 
proportional to $\dot a$. According to our previous discussion, 
\ba \lefteqn{\langle \pi^a (x) \dot\pi^a 
(x)\rangle=3i\lim_{t'\rightarrow 
  t}\int\frac{d^3\vec{k}}{(2\pi)^3}\frac{1}{2} 
\left[\frac{d}{dt}+\frac{d}{dt'}\right] 
G_0^>(t,t',k)}&&\nonumber\\ 
&=&\frac{3}{2}\int\frac{d^3\vec{k}}{(2\pi)^3} \left[1+2 n_B 
  (k)\right]\frac{d}{dt} \vert h_1(t,k\vert^2=\frac{3i}{2}\dot G_0(t) 
\ea where we have used  (\ref{losol}) and (\ref{g0tk}). The above 
result deserves some comments: first, we realize that if we had 
not taken the symmetric limit, the answer would have been complex 
in general, whereas it is manifestly real when the two 
contributions  are added together. This is a consistency check of 
our point-splitting prescription. Second, since 
$h_1(t,k)=i\exp(-ikt)/\sqrt{2k}+\Od(q)$ (the leading order is the 
equilibrium solution) then $\langle \pi^a \dot\pi^a 
\rangle=\Od(q)$. In other words, for the time scales we are 
considering (below $t_{BR}$) $G_0(t)/\fpi^2=\Od(q)$. Therefore and since 
$\dot a=\Od(q)$ as well, to leading order  we just have 
\ba \frac{\langle  E^{(2,2)}(t)\rangle}{V}&=&\frac{3}{2} 
\int\frac{d^3\vec{k}}{(2\pi)^3} \left\{\left[1+2n_B(k)\right] 
\right.\nonumber\\&\times& \left.\left[\vert \dot h_1 
(t,k)\vert^2+ k^2\vert h_1 (t,k)\vert^2\right]\right\} + \Od (q^2) 
\ea

Hence, according to (\ref{nepndef}), we find that the number of 
massless pions of a given type  is given to leading order by 
\ba n(k,t)&=&\frac{1}{2k}\left[\vert \dot h_1 (t,k)\vert^2+ 
k^2\vert h_1 (t,k)\vert^2\right] 
\left[1+2n_B(k)\right]-\frac{1}{2}\nonumber\\\dla{nplo}\ea 
 
This expression coincides with the one given in 
\cite{boyaetal96,linde} in the canonical operator formalism. 
 Note that to this order the answer for the adiabatic number of pions 
 would have been the same, since $m^2(t)=\Od(q)$. 
Note also that our initial conditions (\ref{twoeqsol}) imply that 
$n(k,0)=n_B(k)$ as it should. 
 
So far we have considered the general result in ChPT, for arbitrary 
$\ft$.  Let us now 
particularize to the parametric resonance case in (\ref{ft}). 
First, in order to 
analyse the UV behaviour of  the pion density in 
(\ref{totpndef}), we need the solution $h_1$ to the Mathieu 
equation  in the stable band (large $k$). Using the approximate 
solutions to leading order in $q$ (see Appendix \ref{solmat}) one has: 
\ba \vert h_1(t,k)\vert^2&=&\frac{1}{2k}\left\{ 1+\frac{q 
 M^2}{4} \frac{\sin Mt - \frac{M}{2k}\sin 2kt}{k^2-M^2/4} 
 +\Od(q^2)\right\}\nonumber\\ \vert \dot 
h_1(t,k)\vert^2&=&\frac{k}{2}\left\{ 1-\frac{q 
 M^2}{4} \frac{\sin Mt-\frac{M}{2k}\sin 2kt}{k^2-M^2/4} +\Od(q^2)\right\} 
\nonumber \ea for 
$k>k_0$ (see section \ref{parres}). Thus, we see that the possible 
logarithmic divergence in $d\rightarrow 4$ (i.e, that coming from 
the term which is not proportional to $n_B(k)$ in (\ref{nplo})) 
cancels exactly to leading order in $q$ and therefore the total 
number is finite. This justifies our approach of point-splitting 
the fields as long as we remain within the DR scheme. 
 
The pion number will grow exponentially in time due to the 
contribution of the unstable band. By the same arguments as those 
used for the two-point function, $n(k,t)$  will grow in 
time typically with  $\exp(qMt/2)$. In 
Figure \ref{np} we have plotted $n(k,t)$ for the choice of 
parameters (d) in Figure \ref{unst}. We observe that the pion 
number grows to order one within the resonance band and before the 
back-reaction time. 
 We have also plotted the time average of the pion 
 distribution function $n(k,t)$  in Figure \ref{fig:npav}, taking 
 $Mt_f=30$ (see our comments in section \ref{numres}). 
 
We observe the 
 typical peak of parametric resonance amplification at $k\simeq$ 300 
 MeV, which would be seen in the  final pion spectra \cite{randrup00} 
although  single pion distributions  might be 
 not enough to disentangle from states which are not 
 DCC-like and  one needs to consider higher order 
 pion correlation functions \cite{hiro,ble00}. Note that the final 
 shape is very different from a thermal Bose-Einstein distribution 
 function, although  both  diverge for $k\rightarrow 0$, which  reflects 
 Bose-Einstein condensation for massless particles. 
 The physical pion 
 number density in momentum space is 
 $k^2 n(k,t)$, whose time average we have also plotted in 
 Figure \ref{fig:npav} (dashed line). 
 
\begin{figure} 
\centerline{\epsfig{file=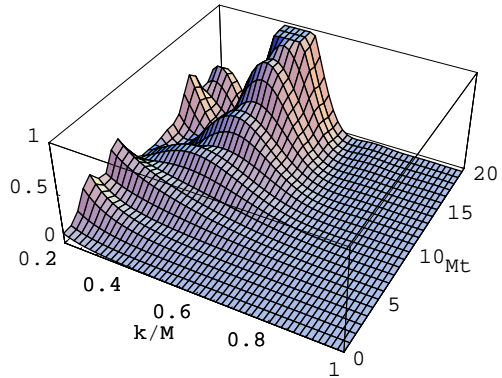,width=8cm}} 
\caption{$n(k,t)$  for $M=600$ MeV, $T_i=50$ 
MeV, $q=0.2$. } \dla{np} 
\end{figure}

\begin{figure} 
\centerline{\epsfig{file=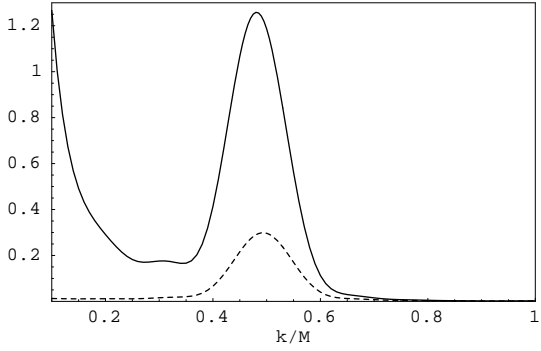,width=8cm}} 
\caption{The solid line  is the averaged pion 
number $\bar n (k,t_f)$ for $t_f=30/M$ and the dashed line is 
$(k^2/M^2) \bar n (k,t_f)$. Here,  $T_i=50$ MeV, $q=0.2$ and 
$M=600$ MeV.}  \dla{fig:npav} 
\end{figure}

 Finally, 
 the pion density (\ref{totpndef}) as a function of time 
 is showed in Figure \ref{fig:density}. 
 Our results agree numerically with the predictions of the $O(4)$ 
 model in the spinodal regime \cite{cooper9596,bodeho95}. 
 As we have previously commented, the particle 
 number grows in time without stop. 
  One may wonder whether the one-loop NLO effects 
  may change this picture, but  this is not the case, 
 as we are going to see in the next section. The origin of this 
 behaviour, as emphasized before, 
  is that we have not taken into account the back-reaction 
  properly  \cite{boyaetal96}. As long as we 
  remain below the back-reaction time, our results for the pion number can be 
  trusted. 
 
\begin{figure} 
\centerline{\epsfig{file=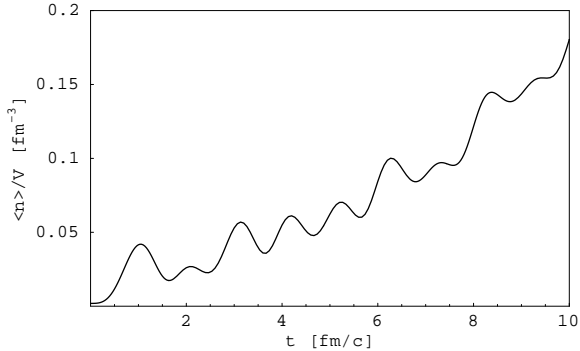,width=8cm}} 
\caption{The pion density for $T_i=50$ MeV, $q=0.2$ 
and $M=600$ MeV.} \dla{fig:density} 
\end{figure}

\subsubsection{Next to leading order} 
 
There are three types of NLO contributions to the pion number: 
 
\begin{enumerate} 
 
  \item \dla{itemloop} The one-loop corrections to the two-point function, 
  which we have already discussed in section \ref{sec:nepddf} 
  (diagrams a) and b) in Figure \ref{diagrams}). 
 
  \item \dla{item42} The contribution of the $\Od(p^4)$ 
lagrangian (\ref{lag4gen}) to the 
  energy-momentum tensor. According to our 
  previous discussion, only two-field terms  contribute to 
  NLO, namely $\langle E^{(4,2)} \rangle$ in our notation. 
 
  \item \dla{item24} Four-field terms in the $\Od(p^2)$ lagrangian, i.e, $\langle E^{(2,4)} 
  \rangle$. 
 
\end{enumerate} 
 
 Before proceeding, we should insist  that, as commented before, 
 we are considering the number of 
   massless pions,  ignoring 
 possible one-loop modifications of the dispersion law. 
 We must stress that the  difficulties associated to the 
   definition of the 
   time-dependent dispersion law are similar to those related to 
the nonequilibrium particle number. Physically it makes sense 
   to define it asymptotically  at long-times 
but it is not clear whether one can actually define an 
   instantaneous dispersion law, unless one follows the adiabatic 
   approximation. 
 Nonetheless, we believe 
   that such corrections would  not change qualitatively the time 
   evolution of the pion number. 
   In  section \ref{fpi} we have already commented on 
the in-medium dispersion law and its relationship with the pion 
decay constants $\fpi^{s,t}$. Remember that in equilibrium the 
pion dispersion law remains unchanged to one-loop in ChPT. 
   While a detailed nonequilibrium extension of the results in 
   \cite{pity96} is out of the scope of this work, qualitatively we expect a 
   similar one-loop behaviour in our case with perhaps 
  a change in the pion velocity of order $\vert \fpisp 
    (t)/\fpite(t) \vert$ 
 as (\ref{timedepvel}) shows.  That contribution is 
 proportional to $L_{12}$ and bounded 
    in time. In any case, if  we write 
   $\omega=k+\Sigma(k,t)$ with $\Sigma(k,t)=\Od(q)$ a bounded function in 
   time,  the  contribution to the pion number (\ref{nepndef}) 
   would be 
    $n\rightarrow n-(\Sigma(k,t)/k)n_B(k)$. This is a bounded 
    correction that does not change the relevant features of pion 
    production we are analyzing here.

Let us then start with the NLO corrections of type \ref{itemloop}. 
According to our power counting, this correction affects only the 
terms in $E^{(2,2)}$. The NLO correction to the pion two-point 
function has been calculated in \cite{agg99} as explained before. Let 
us write it as 
$G^> (t,t',k)=G_0^> (t,t',k)+\Delta(t,t',k)$. Then, according to 
our arguments in the previous section, the contribution  of this 
correction to the pion number is given to $\Od (q)$  by: 
 
$$ n(k,t)\rightarrow n(k,t)+ 
\frac{i}{2k} 
\left[\left.\frac{d}{dt}\frac{d}{dt'}\Delta(t,t',k)\right\vert_{t=t'} 
+ k^2\Delta(t,t,k)\right]$$

Now, from the results in \cite{agg99} (eqs (15)-(17) in that paper 
\footnote{There are two misprints in eq.(16) in \cite{agg99}. The 
term $6 \ddot f(\tilde t )/f(\tilde t)$ multiplying $G_0(\tilde 
t)$ should read $4 \ddot f(\tilde t )/f(\tilde t)$ and the term 
$-2\ddot G_0(\tilde t)$ should read $-\ddot G_0(\tilde t)$. None 
of them affects the results here, since they are included in the 
$\Delta_1 (t,k)$ function in (\ref{prevtype1}).}) we obtain: \ba 
&&i\left[\left.\partial_t\partial_{t'}\Delta(t,t',k)\right\vert_{t=t'} 
+ k^2\Delta(t,t,k)\right]\nonumber\\ &=&-if_1(t) k^2 
G_0^>(t,t)-\frac{i}{2}\dot 
f_1(t)\left[\left.\partial_t+\partial_{t'}\right] 
G_0^>(t,t')\right\vert_{t=t'}\nonumber\\ &+&\left[2\Delta_2 
(t)-if_1 (t)\right] \left.\partial_t\partial_{t'} 
G_0^>(t,t')\right\vert_{t=t'} 
  \nonumber\\ 
&+&\int_0^t du \Delta_1(u,k)\left\{ 
\left[\left(\partial_tG_0^>(t,u)\right)^2 
-\left(\partial_tG_0^>(u,t)\right)^2\right]\right.\nonumber\\ 
&+&\left.k^2\left[\left(G_0^>(t,u)\right)^2 
-\left(G_0^>(u,t)\right)^2\right]\right\} \nonumber\\ &+&\int_0^t 
du \Delta_2(u)\left\{ 
\left[\left(\partial_u\partial_tG_0^>(t,u)\right)^2- 
\left(\partial_u\partial_tG_0^>(u,t)\right)^2\right]\right.\nonumber\\ 
&+&\left.k^2 \left[\left(\partial_uG_0^>(t,u)\right)^2- 
\left(\partial_uG_0^>(u,t)\right)^2\right] \right\} 
\dla{prevtype1} 
 \ea 
 
Here, the $k$-dependence of $G_0^>(t,t',k)$ has been suppressed 
for simplicity, $f_1(t)$ is the function in (\ref{f12}) 
appearing in  the $\Od(p^4)$ lagrangian 
 and we have 
used that $\partial_t[G_0^> (t,t')-G_0^< (t,t')]=-1$. The explicit 
form of the functions $\Delta_1(t,k)$ and $\Delta_2(t)$ is given 
in \cite{agg99}. What is important for our purposes here is that 
$\Delta_{1,2}=\Od(q)$ within the range of validity of our 
approximation, i.e, while $G_0(t)/\fpi^2=\Od(q)$. In fact, 
$\Delta_2(t)=G_0(t)/f^2(t)$, while the precise form of $\Delta_1$ 
is unimportant here (see below). Hence, to $\Od(q)$ it is enough 
to replace in (\ref{prevtype1}) the leading order term in $q$ for 
$G_0^>$, which is nothing but the equilibrium solution 
(\ref{g0+eq}). \footnote{This holds also in the unstable band 
without expanding in $q$ in  the leading exponentials $\exp{\mu 
Mt}$ where $\mu=\Od(q)$ is the Floquet characteristic exponent 
(see Appendix \ref{solmat}).} This simplifies considerably the 
above expression. In fact, the dependence with $\Delta_1(t,k)$ 
disappears and (\ref{prevtype1}) reduces to: \ba 
i\left[\left.\frac{d}{dt}\frac{d}{dt'}\Delta(t,t',k)\right\vert_{t=t'} 
+ k^2\Delta(t,t,k)\right]\nonumber\\ =-k 
\left[1+2n_B(k)\right]\left[f_1(t)+i\frac{G_0(t)}{\fpi^2}\right]+\Od(q^2) 
\dla{pretype1} \ea 
 
 Therefore, 
the correction to the pion number from this part is, neglecting 
$\Od(q^2)$, 
\be 
n(k,t)\rightarrow n(k,t) -\frac{1}{2} 
\left[1+2n_B(k)\right]\left[f_1(t)+i\frac{G_0(t)}{\fpi^2}\right] 
\dla{type1}\ee where 
$f_1(t)=-12\left(2L_{11}+L_{12}\right)\fddot/\fpi^3+\Od(q^2)$.

  Note that 
 in our one-loop calculation of $\fpi$ in section \ref{fpi} we 
 have shown that the combination $-12L_{11}\ddot f/f-iG_0(t)$ is 
 finite. This is the combination appearing in 
 (\ref{fpispnlo})-(\ref{fpitenlo}) with $f_1(t)$ and $f_2(t)$ in 
 (\ref{f12}). However we have  here $-24L_{11}\ddot 
 f/f+iG_0(t)$, which diverges. The only possible way out is 
 then that the remaining NLO corrections (types \ref{item42} and 
 \ref{item24}) 
 combine with this 
 one in such a way that the answer for the pion number is finite. 
 
Let us then consider type \ref{item42} corrections, i.e, those 
coming from the energy-momentum tensor to fourth order in 
derivatives. By the same argument as before, only two-field terms 
contribute to this order. Hence, from the lagrangian in 
(\ref{lag4gen}) we can concentrate only in the $L_{11}$ and 
$L_{12}$ terms. The energy-momentum tensor coming from that 
lagrangian is calculated in Appendix \ref{ap:curved}. The final 
expression for $T_{\mu\nu}^{(4,2)}$ is displayed  in  (\ref{t42}) 
for arbitrary metric and the contribution to the energy for the RW 
metric is given  in (\ref{e42}). Note that we could have used the 
equations of motion to second order to simplify some of the terms 
in  (\ref{e42}), writing for instance $\ddot\pi$ in terms of 
$\Delta\pi$ and $\pi^2$ and so on. 
 It is clear that to $\Od(q)$ 
we only need to consider the terms proportional to $g_1$, $g_5$ 
and $g_7$ in that expression. We have \ba 
\lefteqn{\langle\partial_i\pi^a\vxt\partial_i\dot\pi^a\vxt 
\rangle}&&\nonumber\\ &=&\frac{3i}{2}\lim_{t'\rightarrow t} \int 
\frac{d^3\vec{k}}{(2\pi)^3}k^2\left[\frac{d}{dt}+\frac{d}{dt'}\right] 
G_0^> (t,t',k) \dla{sobra1}\ea and \ba 
\lefteqn{\langle\dot\pi^a\vxt\ddot\pi^a\vxt 
\rangle}&&\nonumber\\&=&\frac{3i}{2}\lim_{t'\rightarrow t} \int 
\frac{d^3\vec{k}}{(2\pi)^3}\frac{d}{dt}\frac{d}{dt'} 
\left[\frac{d}{dt}+\frac{d}{dt'}\right] G_0^> (t,t',k) 
\dla{sobra2}\ea where we have retained only the LO propagator 
$G_0^>$ since the two terms above are multiplied by $g_5 (t)$ and 
$g_7 (t)$ which are already $\Od(q)$. Now, since 
$G_0^>(t,t',k)=G_0^{>eq}(t-t',k)+\Od(q)$ with $G_0^{>eq}(t-t',k)$ 
in (\ref{g0+eq})  it is clear that the contributions 
(\ref{sobra1})-(\ref{sobra2}) are both $\Od(q)$ and therefore they 
do not contribute to the order we are considering here. 
 
Hence, the only correction of this kind is the one given by the 
term proportional to $g_1(t)$ in (\ref{e42}). This term is of 
the form already analyzed in section \ref{sec:numberlo}. From our 
results there, we have that the correction to the particle number 
from $\langle E^{(4,2)}\rangle$ is then given to $\Od (q)$ by 
\be 
n(k,t)\rightarrow 
n(k,t)+\left[1+2n_B(k)\right]\frac{2g_1(t)}{\fpi^3} \dla{type2} 
\ee where $ g_1(t)=-3 (2L_{11}+L_{12})\ddot \ft+\Od(q^2)$.

Finally, we shall consider the contribution of four fields  from 
the $\Od(p^2)$ lagrangian in (\ref{Scurv}) to the energy momentum 
tensor. Expanding to $\Od(\tilde\pi^4)$ in that lagrangian we find 
: 
\be 
T_{00}^{(2,4)}=\frac{f^2(t)}{2f^4} \sum_{a,b=1}^3 
\tilde\pi_a\tilde\pi_b\left[\dot{\tilde\pi_a}\dot{\tilde\pi_b} 
+\left(\nabla\tilde\pi_a\right)\left(\nabla\tilde\pi_b\right)\right] 
\dla{t24} 
\ee

Now we should write the energy (\ref{endef}) to this order for the 
$\pi^a=\tilde\pi^a \ft/f$ fields, from the above expression. 
However, according to our power counting, since this contribution 
to the energy is $\Od(p^2)$ with respect to the tree level, it 
must be at least $\Od(q)$. This will be confirmed by our 
subsequent calculation. Therefore, when changing from the 
$\tilde\pi$ fields to the $\pi$ fields we can simply ignore the 
terms proportional to $\dot \ft$. Hence, to this order it is 
enough to replace simply $\tilde\pi$ by $\pi$ and $\ft$ by $f$ in 
(\ref{t24}) and therefore \ba 
 E^{(2,4)}&=&\frac{1}{2f^2}\int d^3\vec{x}\left\{ 
 \sum_{a,b=1}^3 \pi_a\vxt\pi_b\vxt\left[\dot\pi_a\vxt\dot\pi_b\vxt 
 \right.\right.\nonumber\\ 
&+&\left.\left.\left(\nabla\pi_a\vxt\right) 
\left(\nabla\pi_b\vxt\right)\right]\right\} 
\dla{e24} \ea 
 
We have to calculate now the expectation value of the above 
quantity. Note that this is the first time where we find the 
problem of the $T$-ordering  discussed in section 
\ref{sec:parnum}, apart from the issue of the symmetrization of 
the classical fields. We will follow the prescription explained in 
that section for the different field structures appearing in 
(\ref{e24}). Let us consider first the terms  $\langle 
\pi_a\pi_b\partial_\alpha\pi_a\partial_\alpha\pi_b\rangle$ with 
$a\neq b$ and where the Lorenz index $\alpha$ is fixed, i.e, is 
not summed over. We have \ba \lefteqn{\langle \pi_a (x) \pi_b (x) 
\partial_\alpha\pi_a (x)\partial_\alpha\pi_b (x)\rangle}&&\nonumber\\ 
&=& \frac{1}{4!}\lim^*_{x_j\rightarrow 
x}\left[\partial^{x_3}_\alpha\partial^{x_4}_\alpha \langle T \pi_a 
(x_1) \pi_b (x_2) \pi_a (x_3) \pi_b (x_4)\rangle + 
\dots\right]\nonumber\ea where the dots stand for  all the 
permutations of the four fields in the expectation value. Now, 
using Wick's theorem, \ba \lefteqn{\langle T \pi_a (x_1) \pi_b 
(x_2) \pi_a (x_3) \pi_b (x_4)\rangle}&&\nonumber\\ 
&=&-G_0(x_1,x_3) G_0 (x_2,x_4) 
\left[1+\Od(p^2/\Lambda_{\chi}^2)\right]\quad \mbox{for} \quad 
a\neq b\nonumber\ea which is the only order we have to retain 
according to our power counting (remember that this contribution 
to the energy is already NLO and therefore it is enough to keep 
the LO propagator) and we have taken into account that the lowest 
order action in (\ref{act2pi}) is diagonal in isospin space, so 
that to lowest order the $T$-product of fields with different 
indices factorizes. Note also that the result to this order is 
independent of the $a$, $b$ indices as long as $a\neq b$. 
 Collecting the different permutations  we find 
\ba \lefteqn{\langle \pi_a (x) \pi_b (x) 
\partial_\alpha\pi_a (x)\partial_\alpha\pi_b (x)\rangle}&&\nonumber\\ 
&=& 
-\left[\frac{1}{2}\left.\left(\partial_\alpha^{x}+\partial_\alpha^{x'}\right) 
G_0^> (x,x')\right\vert_{x=x'}\right]^2 \qquad (a\neq 
b)\nonumber\ea 
 
It is clear, following the same arguments as  before, that the 
above quantity is $\Od(q^2)$ since 
$\left(\partial_\alpha^{x}+\partial_\alpha^{x'}\right) G_0^> 
(x,x')=\Od(q)$. Thus, we only need to consider the terms with 
$a=b$ in (\ref{e24}) to $\Od (q)$. For a given $\pi$ field and 
$\alpha$ fixed, \ba \langle \pi^2 (x) \left[\partial_\alpha\pi 
(x)\right]^2\rangle = \frac{1}{6}\lim^*_{x_j\rightarrow 
x}\left(\partial^{x_1}_\alpha\partial^{x_2}_\alpha\right.\nonumber\\ 
+\left.\partial^{x_1}_\alpha\partial^{x_3}_\alpha+\dots\right) 
\langle T \pi (x_1) \pi(x_2) \pi (x_3) \pi (x_4)\rangle. 
\nonumber\ea Thus, Wick's theorem gives now \ba \langle T \pi 
(x_1) \pi(x_2) \pi (x_3) \pi 
(x_4)\rangle=-\left[G_0(x_1,x_2)G_0(x_3,x_4)\right.\nonumber\\ 
+\left.G_0(x_1,x_3)G_0(x_2,x_4)+ G_0(x_1,x_4)G_0(x_2,x_3)\right] 
\nonumber\ea 
 
Now, we take the Fourier transform in the spatial components of 
the above expression. Our prescription for the 
$\lim^*{x_j\rightarrow x}$ is equivalent to replace all the $G_0$ 
above by $G_0^>$ (see our comments in section \ref{sec:parnum}). 
Thus, taking into account once more that 
$G_0^>(t,t',k)=G_0^{>eq}(t-t',k)+\Od(q)$, 
$G_0^>(t,t,k)=G_0(t,t,k)$ and  $G_0(t)/f^2=\Od(q)$,  we find for 
the combination appearing in (\ref{e24}), \ba \langle \pi^2 \vxt 
\left\{\left[\dot\pi \vxt\right]^2 + 
 \left[\nabla\pi 
\vxt\right]^2\right\}\rangle\nonumber\\ 
= 
 iG_0 (t)\int \frac{d^{d-1}\vec{k}}{(2\pi)^{d-1}} 
k\left[1+2n_B(k)\right]+\Od(q^2) 
\dla{nueva} 
\ea

Therefore, from (\ref{e24}) we find that the contribution of the 
corrections of type \ref{item24} to the particle number is given 
by 
\be 
n(k,t)\rightarrow n(k,t)+\left[1+2n_B(k)\right]\frac{i 
G_0(t)}{2\fpi^2} \dla{type3} \ee

Finally, collecting the contributions  to the pion number to NLO, 
namely, (\ref{type1}), (\ref{type2}) and (\ref{type3}) we find an 
interesting result: the total NLO correction to $\Od(q)$ {\it 
vanishes}. Note the completely different origin of these three 
contributions and remember that each of them was UV divergent so 
that there were only two alternatives: either  they cancel or they 
appear in the combination 
 (\ref{fpispnlo})-(\ref{fpitenlo}). Thus, the only  NLO 
 correction to the pion number is due to the change in the 
 dispersion law. Note that, in turn, we have shown the 
 absence of NLO corrections to the pion number for 
 equilibrium ChPT. This is indeed consistent since we know that in 
 equilibrium in the chiral limit the pion dispersion law is unchanged 
 to one loop in ChPT. Therefore, the pion distribution function has to 
 be the Bose-Einstein one for massless particles since the system 
 remains in thermal equilibrium.

 Thus, the numerical results showed 
 in Figures \ref{np} and \ref{fig:npav} 
  remain valid to NLO. As commented before, we did 
 not expect that the NLO corrections make the pion number stop 
 growing since we have not included the back reaction and 
the energy $E(t)$ to this order is 
 still not conserved. Our results are valid below the back reaction 
 time and should account for all the relevant pion production.

\section{Conclusions and Outlook} 
\dla{conc}

Chiral perturbation Theory can be used to describe nonequilibrium 
phenomena. In particular, in this work we have showed that pion 
production can be accommodated in ChPT in the parametric resonance 
regime. The physical situation where this analysis is meant to be 
useful is the late time expansion of the plasma formed after a 
Relativistic Heavy Ion Collision. Pion production is important in 
the context of   hadronization and production of Disoriented 
Chiral Condensates during the chiral phase transition. 
 
In the present approach, we have considered  the Nonlinear Sigma 
Model in the chiral limit, where the pion decay constant is 
time-dependent. This is   a nonequilibrium effective 
 model with a well-defined perturbative expansion and power counting 
near equilibrium. Besides, using the analogy of this model with 
curved space-time QFT, we have been able 
 to construct the  fourth order lagrangian and 
implement renormalization in a consistent fashion. The parametric 
resonance regime corresponds to take $\fpi (t)$ oscillating around 
its equilibrium position. To lowest order  $\fpi (t)$ corresponds 
to the vacuum expectation value of the $\sigma$ field in the 
$O(4)$ model. Thus, 
 the pion equation of motion to lowest order in the 
amplitude oscillations becomes a Mathieu equation, which has 
resonance bands in momentum space. The pion correlator grows 
exponentially in time, yielding  explosive pion production. This 
approximation is consistent until the time when the back reaction 
effects due to the pion correlations become important. We have 
estimated this time scale for different choices of the initial 
values of the amplitude, frequency and temperature. 
 
The two observables we have analyzed here are 
 the pion decay constants  and 
 the pion number up to one loop in ChPT. Our main results are the 
 following. 
For $\fpi (t)$  the 
nonequilibrium corrections are  basically of 
oscillatory nature until the back reaction time. 
However, the central value tends to decrease, which can be interpreted 
in terms of a reheating of the system. Besides, 
 a small difference between $\fpisp$ 
and $\fpite$ is induced, unlike the equilibrium case where it 
vanishes at one loop. Using the equilibrium result, we have 
estimated  the final temperature and  the averaged value of 
$\fpisp-\fpite$ which is related to the in-medium pion velocity in 
equilibrium.

As for the particle number, we have first introduced a suitable 
definition in terms of the energy-momentum tensor and Green 
functions. We have showed that in dimensional regularization there 
is no need for extra renormalizations and we can use a 
point-splitting prescription consistently. The number of initial 
particles at tree level coincides basically with the result in the 
$O(4)$ model. At one loop we have found that all the relevant 
contributions cancel, to leading order in the oscillations 
amplitude. 
 This result holds also in equilibrium, which is a particular case of 
 this analysis. 
Thus,  our prediction for the particle number is just the tree level 
 result, which gives pion exponential growth in time. We have given 
 numerical results both for the pion  distribution function 
 $n(k,t)$ 
 and for the 
 pion density $\langle n(t)\rangle/V$. The final distribution function 
 which would be observed has the typical peak of parametric resonance 
 at the center of the unstable band ($k\simeq M/2$).

The reason why  $\fpi (t)$ and $\langle n(t)\rangle/V$ are not 
 damped is because we have not taken into 
 account the back-reaction effects which would change the original 
 ansatz for the pion decay constant. When taken into account, those 
 effects should make the particle number stop growing and give energy 
 conservation. Nevertheless, our approach is 
 perfectly valid until the time where these dissipation effects are 
 important and therefore we believe we capture  the essential 
 behaviour concerning explosive pion production.

We must stress than, apart from being a physically interesting 
case, the example analyzed here has allowed us to show 
explicitly the renormalization of our model to one loop, which is 
not trivial because of the presence  of new nonequilibrium 
infinities. In fact, we believe that our methods could be useful for 
other nonequilibrium field theoretical models.

There are many directions in which this work can be extended. 
Perhaps the most important would be to be able to include the 
above mentioned  back-reaction effects, in order to understand 
dissipation properly. Other extensions  include to consider 
nonzero  physical pion masses and  other relevant observables such 
as 
 the correlation length or 
  higher pion correlators, which are  important to clarify the issue of  DCC 
 formation and to obtain predictions testable in RHIC. In addition, 
 photon production can be studied by gauging the NLSM, including the 
 Wess-Zumino-Witten term, responsible for the anomalous decay 
$\pi^0\rightarrow\gamma\gamma$. 
 Work along these lines is in progress.

\acknowledgements 
 The author wishes to thank  A.L.Maroto  for  useful 
 comments and discussions. 
Financial 
 support from CICYT, Spain,  projects AEN97-1693 and FPA2000-0956 are 
 acknowledged. 
 
\appendix 
 
\section{Solutions of the  Mathieu equation} 
\dla{solmat} 
 
 Here we will summarize the main results used in the text 
concerning the solutions of  Mathieu differential equation. 
All 
these results can be found in \cite{mac,abramo}. 
 
 According to Floquet's 
Theorem, there is always  a solution of the Mathieu equation 
(\ref{mathieu}) of the form 
\be 
F_\nu (z)=e^{i\nu  z} P(z) \dla{floquet}\ee where $P(z)$ is a 
periodic function with period $\pi$ and $\nu$ is called the 
characteristic exponent, which depends on $a$ and $q$ and it plays 
a crucial role in our analysis, since it gives rise to 
exponentially growing solutions whenever it takes complex values. 
 
The values of $a$ such that $F_\nu (z)$ is periodic in $z$ are 
called the eigenvalues of the Mathieu equation. They correspond to 
integer values of $\nu$. They are denoted as $a_r (q)$ if $\nu$ is 
a positive integer $r$, and $b_r (q)$ if $\nu=-r$. It can be shown 
\cite{mac}  that for $a>0$ one has $a_r>b_r$, $\nu$ is complex in 
the bands $b_r<a<a_r$ and real elsewhere. Moreover, for small $q$, 
one has $b_r-a_r=\Od(q^r)$. Therefore, in the narrow resonance 
regime we are considering here, we will take $\nu$ complex for 
$b_1<a<a_1$ and $\nu$ real for $0<a\leq b_1$ and $a\geq a_1$. The 
series expansion in $q$ of the eigenvalues is given by 
\be 
a_1=1+q+\Od (q^2) \quad ; \quad b_1 =1-q+\Od (q^2)  \ee 
 
If $a\neq a_r,b_r$ then $F_\nu (-z)$ is a solution linearly 
independent of $F_\nu (z)$. This is no longer true if $\nu$ is an 
integer, although an independent solution can also be constructed 
in that case \cite{abramo}. 
For $\nu^2\neq r^2$, it is customary to take as independent 
solutions: 
\ba ce_\nu (z)&=&\frac{1}{2}\left[ F_\nu (z) + F_\nu (-z) 
\right]\nonumber\\se_\nu (z)&=&\frac{1}{2i}\left[ F_\nu (z) - 
F_\nu (-z) \right] 
\dla{ceseF} 
 \ea 
 
For $\nu=r$, $ce_r (z)$ are called the eigenfunctions of the 
Mathieu equation and so on for $\nu=-r$ and $se_r (z)$. They are 
$2\pi$-periodic and their $q$-expansion for $\nu^2=1$ is given by 
\cite{abramo}: 
\be ce_1 (z)=\cos z - \frac{q}{8}\cos (3z) \quad ; \quad se_1 
(z)=\sin z - \frac{q}{8}\sin (3z) \dla{ceseasy}\ee

Therefore, the solution $h_1 (z,k)$ of (\ref{mathieu}) is 
given as a linear combination of $ce_\nu$ and $se_\nu$: 
\be 
h_1 (z,k)=A (k) ce_\nu (z,k) + B(k) se_\nu (z,k) \dla{f1cese}\ee 
where the coefficients $A(k)$ and $B(k)$ are such that 
the initial conditions (\ref{ic}) are satisfied, i.e, 
 
\ba 
A (k) ce_\nu (-\pi/4,k) + B(k) se_\nu (-\pi/4,k)&=&\frac{i}{\sqrt{2k}} 
\nonumber\\ 
A (k) \dot ce_\nu (-\pi/4,k) + B(k) \dot 
se_\nu (-\pi/4,k)&=&\frac{\sqrt{2k}}{M} 
\nonumber 
\ea 
where the dot  means $d/dz$.  Remember that $ce_\nu$ and $se_\nu$ 
depend on $k$ through  $a(k)$. 
 
Even though the solutions to the Mathieu differential equation are 
numerically tabulated, we need their explicit form when dealing 
with renormalization.  Such a explicit form of the solutions can 
be found  for small $q$. Let us consider first  the 
case when $\nu^2\neq r^2$ and {\it real}, i.e, the stable zone. 
Then, using the $q$-expansions given in \cite{mac,abramo}, one has 
$\nu=\sqrt{a}+\Od (q^2)$ and the solutions are given by: 
\ba ce_\nu (z)&=&\cos (\sqrt{a} z) + \tilde q \left[ \cos(\sqrt{a} 
z) \cos 2z \right.\nonumber\\ 
&+& \left.\sqrt{a} \sin (\sqrt{a} z) \sin 2z\right]+\Od 
(q^2)\nonumber\\se_\nu (z)&=&\sin (\sqrt{a} z) + \tilde q \left[ 
\sin(\sqrt{a} z) \cos 2z\right. 
\nonumber\\ 
&-&\left. \sqrt{a} \cos (\sqrt{a} z) \sin 
2z\right]+\Od (q^2) \dla{cestab}\ea where 
\be\tilde q=\frac{a-1}{2 (a-1)^2-q^2} \ q \dla{cesqtil}\ee 
 
If $a$ is far enough from the border points of the first band 
(placed at $a\simeq 1\pm 
q$) we can simply take $\tilde q\simeq q/[2(a-1)]$ in 
(\ref{cestab}). The value of $a$ from which this simplification is 
valid can be estimated numerically, for a given $q$, by comparing 
to the numerical (tabulated) solutions and imposing that the 
difference with the approximate solution (\ref{cestab}) remains 
$\Od (q^2)$.

Now, consider the unstable band, i.e, $b_1<a<a_1$. In this case, 
the solution (\ref{floquet}) reads, to leading order in 
$q$ \cite{mac}: 
\be 
F_\mu (z)=e^{\mu z} \left[ C_1 ce_1 (z)+S_1 se_1 (z)\right] 
\dla{Funst} 
 \ee 
where $ce_1 (z)$ and  $se_1 (z)$ are given asymptotically in 
(\ref{ceseasy}) and 
the {\it real} characteristic exponent $\mu$ and the $C_1$, 
$S_1$ coefficients  are given by 
\ba \mu&=&\frac{1}{2} \sqrt{(a_1 - a) (a-b_1)} \nonumber\\ 
C_1&=&\sqrt{a-b_1+\mu^2}\nonumber\\ S_1&=&\sqrt{a_1-a-\mu^2} 
\dla{muunst} \ea 
 
Note that $\mu=\Od (q)$ and it reaches its maximum value at the band 
center. The dominant behaviour at long times is therefore given by the 
positive exponentials when (\ref{Funst}) is replaced in (\ref{ceseF}) 
and (\ref{f1cese}). 
 
\section{Results in curved space-time} 
\dla{ap:curved}

\subsection{General results} 
 
 We 
will  collect here some of the results concerning curved 
space-time needed for our purposes. In this section we will 
consider an arbitrary metric $g_{\mu\nu}$ and in the next one we 
will particularize  for the spatially flat RW metric. Most of the 
definitions used here 
 can be found in any textbook on the subject (we are following the 
 notation and conventions of \cite{weinberg}). 
 
The covariant derivative of a contravariant vector $V^\mu (x)$ 
satisfies 
\be 
\sqrt{-g} V^\mu_{;\mu}=\partial_\mu \left(\sqrt{-g} 
  V^\mu\right)\Rightarrow \int d^4 x \sqrt{-g}  V^\mu_{;\mu}=0 
\dla{gaussgen} \ee which is the generalized Gauss theorem and 
$V^\mu(x)$ is assumed to vanish at the space-time boundary.

The Ricci tensor and scalar of curvature are defined respectively 
as  $R_{\mu\nu}=R_{\mu\lambda\nu}^\lambda$ and $R=g^{\mu\nu} 
R_{\mu\nu}$ where the  Riemann tensor is: 
\be 
R_{\beta\gamma\delta}^\alpha 
=\partial_{\delta}\Gamma_{\beta\gamma}^{\alpha}- 
\partial_{\gamma}\Gamma_{\beta\delta}^{\alpha} 
+\Gamma_{\delta\lambda}^{\alpha}\Gamma_{\beta\gamma}^{\lambda}- 
\Gamma_{\gamma\lambda}^{\alpha}\Gamma_{\beta\delta}^{\lambda}\ee 
and the Christoffel symbols are given in terms of the metric as 
\be 
\Gamma_{\mu\nu}^{\lambda}=\frac{1}{2}g^{\lambda\alpha}\left[\partial_\mu 
g_{\nu\alpha}+ \partial_\nu g_{\mu\alpha}-\partial_\alpha 
g_{\mu\nu}\right] \ee

The classical energy-momentum tensor of the theory $T_{\mu\nu} 
(x)$ is defined by performing a general coordinate transformation 
(infinitesimal)  $g^{\mu\nu}(x)\rightarrow g^{\mu\nu}(x)+\delta 
g^{\mu\nu}(x)$, under which the action $S=\int \sqrt{-g}{\cal L}$ 
changes as $S\rightarrow S+\delta S$ where by definition 
 
$$ \delta S =\frac{1}{2} \int \sqrt{-g} T_{\mu\nu} \delta 
g^{\mu\nu}$$ 
 
The energy-momentum tensor thus defined is symmetric and conserved 
($T_{\mu;\nu}^{\nu}=0$) as long as the action is a Lorenz 
scalar.  Under the above transformation the variation of 
the metric determinant is given by 
\be 
\delta \sqrt{-g}=-\frac{1}{2} \sqrt{-g} g_{\mu\nu} \delta 
g^{\mu\nu}  \dla{vardet} \ee whereas  the variation of the Ricci 
tensor yields \cite{weinberg}: \ba \delta 
R_{\mu\nu}&=&\frac{1}{2}g^{\rho\lambda} \left[\left(\delta 
g_{\rho\lambda}\right)_{;\mu;\nu} - \left(\delta 
g_{\rho\mu}\right)_{;\nu;\lambda}\right.\nonumber\\ &+& 
\left.\left(\delta g_{\mu\nu}\right)_{;\rho;\lambda}- \left(\delta 
g_{\rho\nu}\right)_{;\mu;\lambda}\right] \dla{pala}\ea 
 which is known as the Palatini identity.

In the text we need the energy-momentum tensor defined in 
(\ref{tmunudef}), to fourth order, i.e, when the lagrangian is 
given by (\ref{lag4gen}). Since we are interested only in 
two-point functions, we can ignore the contribution of ${\cal L}_4 
(U,g)$  and concentrate only in the $L_{11}$ 
and $L_{12}$ pieces. Therefore, using (\ref{vardet}), 
\be 
T_{\mu\nu}^{(4,2)} (x)=-g_{\mu\nu} (x){\cal 
L}^{(4,2)}(x)+2\frac{\delta {\cal L}^{(4,2)}(x)}{\delta 
g^{\mu\nu}(x)}\dla{t42l42}\ee with \ba {\cal 
L}^{(4,2)}(x)=-\left[L_{11}R(x) 
g^{\mu\nu}(x)+L_{12}R^{\mu\nu}(x)\right] F_{\mu\nu}(x)\nonumber\\ 
F_{\mu\nu}=\tr\left[ 
\partial_\mu U^\dagger(x) 
\partial_\nu U(x)\right] =\frac{2}{f^2}\partial_\mu\tilde\pi^a(x) 
\partial_\nu\tilde\pi^a(x) 
+ \Od(\tilde\pi^4)\nonumber\ea 
in the parametrization of the 
$\tilde\pi (x)$ fields.

To calculate the variation of the lagrangian in (\ref{t42l42}) we 
use (\ref{pala}) and integrate by parts taking into account 
(\ref{gaussgen}) as well as the properties of the covariant 
derivative.  We find after a straightforward but lengthy 
calculation: \ba T_{\mu\nu}^{(4,2)} 
&=&-L_{11}\left[\left(2R_{\mu\nu}-g_{\mu\nu} 
R\right)g^{\alpha\beta} F_{\alpha\beta} +2RF_{\mu\nu} 
\right.\nonumber\\ &-&\left.2g_{\mu\nu}g^{\alpha\beta} 
{F_{\alpha\beta}}^{;\delta}_{;\delta} + g^{\alpha\beta} 
\left(F_{\alpha\beta;\mu;\nu}+F_{\alpha\beta;\nu;\mu}\right)\right] 
\nonumber\\ &-& L_{12}\left[2\left(R_\mu^\alpha 
F_{\nu\alpha}+R_\nu^\alpha F_{\mu\alpha}\right)-g_{\mu\nu} 
R^{\alpha\beta} F_{\alpha\beta} \right.\nonumber\\ &-&\left. 
g_{\mu\nu} {F_{\alpha\beta}}^{;\beta;\alpha} 
{F_{\mu\nu}}^{;\delta}_{;\delta}+{F_\nu^{\beta}}_{;\mu;\beta} + 
{F_\mu^{\beta}}_{;\nu;\beta} \right] \dla{t42}\ea

The above energy-momentum tensor is symmetric, since $g_{\mu\nu}$, 
$R_{\mu\nu}$ and $F_{\mu\nu}$ are  symmetric. Note that the 
symmetry of $F_{\mu\nu}$  is a consequence of 
$U$ being unitary  so that 
 
$$\tr[\partial_\nu U^\dagger \partial_\mu U]= 
\tr[U \partial_\nu U^\dagger \partial_\mu U U^\dagger]= 
\tr[\partial_\mu U^\dagger \partial_\nu U]$$

On the other hand, in Minkowski space-time, where $R_{\mu\nu}=0$ 
and the covariant derivatives are ordinary partial derivatives, we 
recover the result given in \cite{dole91}. 
 Finally, we have verified explicitly that 
 ${T^{(4,2)}}_{\mu;\nu}^{\nu}=0$, 
using the equations of motion for ${\cal L}^{(4,2)}$, which to 
 $\Od(\tilde\pi^2)$ read 
\be 
\left\{\left[L_{11}R 
g^{\mu\nu}+L_{12}R^{\mu\nu}\right]\partial_{\nu}\tilde\pi\right\}_{;\mu}=0 
\dla{eom42} \ee

\subsection{Results for the RW conformal metric} 
 
As explained in the text, the space-time metric we are 
interested in is the RW spatially flat metric in conformal time, 
where the scale factor is $a(t)=f(t)/f$. The line element is 
$ds^2=a^2(t)[dt^2-d\vec{x}^2]$ so that the elements of the metric 
are: 
 
$$ g_{00} (t)=a^2(t), \ \ g_{ij}(t)=-\delta_{ij} a^2 (t), \ \ 
g_{i0}=0 $$ and the metric determinant is $g\equiv\det g=-a^{8} 
(t)$. The nonvanishing Christoffel symbols for this metric are: 
 
$$ \Gamma_{00}^0 (t)=\frac{\dot a (t)}{a(t)}, \ \ 
\Gamma_{0i}^{k}=\frac{\dot a (t)}{a(t)} \delta_i^k, \ \ 
\Gamma_{ij}^{0}=\frac{\dot a (t)}{a(t)} \delta_{ij}, $$  and the 
 nonvanishing elements of the Ricci tensor are given by: 
 
$$ R_{00} (t)=3\left[\frac{\ddot a (t)}{a(t)}-\frac{\dot a^2 
(t)}{a^2(t)}\right], \ \ R_{ij}= -\left[\frac{\ddot a 
(t)}{a(t)}+\frac{\dot a^2 (t)}{a^2(t)}\right] \delta_{ij}$$ so 
that the scalar of curvature is 
 
$$R(t)=6 \frac{\ddot a (t)}{a^3(t)}.$$

With the above ingredients we can calculate the energy-momentum 
tensor in (\ref{t42}) for this metric. It is easy to check that 
$\langle T_{\mu\nu}^{(4,2)}\rangle$ is diagonal for the RW metric 
above, i.e, $\langle T_{i0}^{(4,2)}\rangle=0$ and $\langle 
T_{ij}^{(4,2)}\rangle=0$ if $i\neq j$, as it happens also with the 
lowest order $\langle T_{\mu\nu}^{(2,2)}\rangle$ (see section 
\ref{sec:numberlo}). 
Therefore, we only need $T_{00}$ 
  to calculate the total energy defined in (\ref{endef}) 
to this order.  We give the result here in terms of 
the $\pi$ fields and the $\ft$ 
 function, retaining only two-field terms: 
 
\ba  \lefteqn{E^{(4,2)} (t) = \int d^3 \vec{x} a^2(t) 
T^{(4,2)}_{00}}&&\nonumber\\ &=&\frac{4}{f^3 (t)}\int d^3 
\vec{x}\left\{ 
g_1(t)\left[\dot\pi^a\right]^2+g_2(t)\left[\nabla\pi^a\right]^2 
+g_3(t)\left[\pi^a\right]^2 \right.\nonumber\\ &+&\left. 
g_4(t)\pi^a\dot\pi^a + g_5(t)(\nabla 
\pi^a)\cdot(\nabla\dot\pi^a)+g_6(t)\pi^a\ddot\pi^a \right.\nonumber\\ 
&+&\left.g_7(t)\dot\pi^a\ddot\pi^a\right\}\dla{e42}\nonumber\\\ea where we 
have integrated by parts neglecting total spatial derivatives, 
 the space-time dependence of the fields has been suppressed 
for simplicity and 
\ba g_1(t)&=&-3 (2L_{11}+L_{12})\ddot \ft-3 
(5L_{11}+L_{12})\frac{\dot \ft^2}{\ft}\nonumber\\ 
g_2(t)&=&(9L_{11}+2L_{12})\frac{\dot \ft^2}{\ft}\nonumber\\ 
g_3(t)&=&-3  (5L_{11}+L_{12})\frac{\dot \ft^4}{f^3(t)}\nonumber\\ 
g_4(t)&=&3\frac{\dot\ft}{\ft}\left[(2L_{11}+L_{12})\ddot\ft 
+2(5L_{11}+L_{12})\frac{\dot \ft^2}{\ft}\right]\nonumber\\ 
g_5(t)&=&-(6L_{11}+L_{12})\dot\ft\nonumber\\ 
g_6(t)&=&-3(2L_{11}+L_{12})\frac{\dot \ft^2}{\ft}\nonumber\\ 
g_7(t)&=&3(2L_{11}+L_{12})\dot\ft \ea

\end{document}